\documentclass[aps,prd,preprint,superscriptaddress,nofootinbib,showpacs]{revtex4}

\pdfoutput=1
\usepackage{graphicx}
\usepackage{hyperref}
\usepackage{ulem}
\usepackage{color}
\usepackage{slashed}
\definecolor{My_red}        {cmyk}{0.00,1.00,1.00,0.20}

\newcommand{\bmat}{\left(\begin{array}}
\newcommand{\emat}{\end{array}\right)}
\newcommand{\beq}{\begin{equation}}
\newcommand{\eeq}{\end{equation}}
\newcommand{\wt}{\widetilde}

\newcommand{\ET}{\mbox{$\not \hspace{-0.10cm} E_T$ }}

\def\ra{\rightarrow}

\def\Ld{\Lambda}
\def\ld{\lambda}
\def\f{\frac}

\def\bwt{\begin{widetext}}
\def\ewt{\end{widetext}}
\def\be{\begin{equation}}
\def\ee{\end{equation}}
\def\bary{\begin{array}}
\def\eary{\end{array}}
\def\bit{\begin{itemize}}
\def\eit{\end{itemize}}

\def\ra{\rightarrow}

\def\Ld{\Lambda}

\def\ld{\lambda}

\def\su5u1{SU(5) \times U(1)}
\def\fsu5u1{SU(5) \times U(1)'}
\def\so10{SO(10)}
\def\sq20{SO(10) \times SO(10)}

\def\ra{\rightarrow}

\def\Ld{\Lambda}
\def\ld{\lambda}
\def\f{\frac}

\def\L{\left(}
\def\R{\right)}
\def\bwt{\begin{widetext}}
\def\ewt{\end{widetext}}
\def\be{\begin{equation}}
\def\ee{\end{equation}}
\def\bary{\begin{array}}
\def\eary{\end{array}}
\def\bit{\begin{itemize}}
\def\eit{\end{itemize}}

\def\ra{\rightarrow}

\def\Ld{\Lambda}

\def\ld{\lambda}

\def\su5u1{SU(5) \times U(1)}
\def\fsu5u1{SU(5) \times U(1)'}
\def\so10{SO(10)}
\def\sq20{SO(10) \times SO(10)}

\usepackage[centertags]{amsmath}
\usepackage{amssymb}

\begin{document}

\title{New  Avenues to Heavy Right-handed Neutrinos with \\
 Pair Production at Hadronic Colliders}


\author{Zhaofeng Kang}
\email[E-mail: ]{zhaofengkang@gmail.com}
\affiliation{School of Physics, Korea Institute for Advanced Study,
Seoul 130-722, Korea}

\author{P. Ko}
\email[E-mail: ]{pko@kias.re.kr}
\affiliation{School of Physics, Korea Institute for Advanced Study,
Seoul 130-722, Korea}

\author{Jinmian Li}
\email[E-mail: ]{phyljm@gmail.com}
\affiliation{ARC Centre of Excellence for Particle Physics at the Terascale, Department of Physics, University of Adelaide, Adelaide, SA 5005, Australia}

\date{\today}

\begin{abstract}

In many models incorporating the type-I seesaw mechanism, the right-handed neutrino ($N$)  couples to heavy vector/scalar bosons and thereby has resonant pair production. It barely receives attention thus far, however, it may provide the best avenue to probe TeV scale $N$ without requiring anomalously large mixing between $N$ and the active neutrino $\nu_L$. In this paper we explore the discovery prospects of (mainly heavy) $N$ pair production at the 14 TeV LHC and future 100 TeV $pp$ collider, based on the three signatures: 1) trilepton from $ N(\ra \ell W_\ell)N(\ra \ell W_h)$ with $W_{\ell/h}$ the leptonically/hadronically decaying $W$; 2) boosted di-Higgs boson plus \ET from $ N(\ra \nu_L h)N(\ra \nu_L h)$; 3) a single boosted Higgs with leptons and \ET from $N(\ra \ell W_\ell)N(\ra \nu_L h)$. At the 100 TeV collider, we also consider the situation when the Higgs boson is over boosted  thus losing its jet substructure. Interpreting our tentative results in the benchmark model, the local $B-L$ model, we find that the (multi-) TeV scale $N$ can be probed at the (100) 14 TeV colliders.


\end{abstract}

\maketitle

\section{Introduction and Motivation}

That neutrinos possess tiny but non-vanishing masses is one of the most confirmative evidence that the standard model (SM) is not a complete theory and we should go beyond it. For instance, one can introduce right-handed neutrinos (RHNs) $N_R$ and realize the canonical seesaw mechanism~\cite{seesaw},
\begin{align}\label{seesaw}
{\cal L}_{seesaw}=-y_N\bar \ell_L \wt H N_R-\f{M_{N}}{2}\overline{ (N_R)^c}  N_R+h.c.,
\end{align}  
where $\wt H=i\sigma_2 H^*$ with $H$ the SM Higgs doublet. For illustration, only one generation is considered here. This canonical seesaw mechanism offers the most elegant and economical explanation to the origins of nonzero neutrino masses. However, its tests at current and future colliders are not that promising for two reasons.

First, RHNs are singlets with respect to SM so they can be produced via neither electroweak nor strong interacting processes. Second, the mass of active neutrino and its mixing with RHN are estimated by 
\begin{align}\label{}
m_\nu\simeq \f{y_N^2v^2}{2M_N}~ (v=246\, {\rm GeV}),\quad V_{iN}\simeq \sqrt{m_\nu/M_N} ~~(i=e, \mu, \tau)
\end{align}  
Thus, the RHN is either extremely heavy and thus not accessible at colliders, or extremely weakly coupled to SM particles, suppressed by $y_N\ll1$ for a weak 
scale RHN. But a nontrivial flavor structure may allow significant deviations from 
the above estimation on mixing angle and a sizable mixing angle can be realized
~\cite{Kersten:2007vk,Huitu:2008gf,Deppisch:2015qwa,He:2009ua,Arganda:2015ija,Das:2012ze,Gluza:2002vs},
also known as low scale seesaw~\cite{Mohapatra:1986bd,Bernabeu:1987gr,Malinsky:2005bi}
Then, it is possible to probe the RHN sector by means of:
\begin{enumerate}
\item The signature containing same-sign dilepton~\footnote{With CP phases and non-degenerate RHN, the opposite-sign dilepton signature can be dominant. It can be used to explain the recent CMS excess~\cite{Gluza:2015goa}. } $pp\ra W^*\ra N_R\ell^\pm \ra \ell^\pm\ell^\pm jj$ with $\ell=e/\mu$~\cite{Datta:1993nm,Hantao,Ng:2015hba}, which is most sensitive to $M_N$ below $M_W$ such that the production cross section is resonantly enhanced. For instance, the CMS 20 fb$^{-1}$ data at $\sqrt{s}=8$ TeV can exclude $|V_{\mu N}|^2\gtrsim3\times 10^{-6}$ for $M_N\lesssim M_W/2$~\cite{CMS}; an improvement is possible after taking into account the contribution from the $N_R\ell j$ production~\cite{Das:2015toa}. The sensitivity  deteriorates quickly for heavier RHN, e.g., in the light of a recent study~\cite{Ng:2015hba}, to probe $M_N=1$ TeV, one has to accumulate 3000 fb$^{-1}$ data at $\sqrt{s}=13$ TeV even if $|V_{e N}|^2$ is as large as $2\times 10^{-2}$~\footnote{The authors also study the search at 100 TeV machine with integrated luminosity of 3000 fb$^{-1}$, and find that the improvement is limited, hardly approaching the region $|V_{\ell N}|^2\lesssim 10^{-3}$ for $M_N>$1 TeV. } and moreover taking into account the photon-mediated production $pp \ra  W^*\gamma^* \ra  N_R\ell^\pm jj $ which benefits in infra-red enhancement~\cite{Dev:2013wba,Alva:2014gxa}.

\item The displaced vertex search at LHC which has negligible SM background is sensitive to light and long-lived RHN from either $W$ boson~\cite{displaced:wdecay,Izaguirre:2015pga,Dib:2015oka} or Higgs decay~\cite{Gago:2015vma}. However, the search turns out to be invalid for heavy RHN.

\item  Searching for channels like $e^+e^-\ra N_R\nu_L,\,N_Re^\pm W^\mp$, and so on~\cite{Antusch:2015mia,Banerjee:2015gca,Antusch:2015gjw} at a lepton collider which has clean environment. 
But the search limits on RHN mass at lepton colliders are bounded by their collision energy, which is typically much lower than that of hadron colliders.

\end{enumerate}
In summary, in order to probe a RHN with mass at least a few hundred GeVs, one needs a sizable active-sterile neutrino mixing angle~\cite{Antusch:2014woa}, which definitely has been excluded by the indirect constraints like Electroweak Precision Tests (EWPT). Moreover, the chance opens only for the mixing with light lepton flavors. In other words, the search will be highly  dependent of flavor models. Therefore, it is justified to conclude that there is very little chance to probe TeV scale RHN in the simplified framework, Eq.~(\ref{seesaw}).

However, those RHNs could have additional interactions which allow an abundant  production of RHNs even in the decoupling limit between the RHN and active neutrinos. A good example is the local $B-L$ extended SM models (BLSM)~\cite{Mohapatra:1980qe,Wetterich:1981bx,Khalil:2006yi} where the RHN pair couples to both new heavy vector and scalar boson $X$ that breaks $U(1)_{B-L}$ gauge symmetry spontaneously. The bosons can mediate the RHN pair production in the $s-$channel~\cite{Khalil:2010iu,Dib:2014fua}, admitting a resonant enhancement. In this paper we concentrate on searching for RHNs in pair production, which offers new avenues to probe RHNs, in particular in the heavy RHN region that is hardly accessible via the conventional search strategies summarized above. 

We study three channels, $WW$, $hh$ (with $h$ the SM-Higgs boson) and as well $hW$, in detail, both at the 14 TeV LHC and at the future 100 TeV $pp$-collider. We attempt to draw a tentative global picture of RHN pair searches on the $M_N-M_{X}$ plane with $M_X\gtrsim 2M_N$. In most of the parameter space on this plane, the Higgs boson  from heavy RHN decay is expected to be highly (even over) boosted  and therefore even the pure hadronic channel can be searched for, says  boosted di-Higgs boson plus \ET~\cite{Kang:2015nga}. 
Typically,  the $WW$ channel is the most hopeful. But in certain parameter space the mixed channel $hW$ or even the $hh$ channel instead can provide the strongest sensitivity. We apply our searches to the benchmark model BLSM and find that the multi-TeV RHN can be probed at 14 TeV LHC; at the 100 TeV collider, the remarkable 10 TeV mass scale is possible, which enables us to cover most of the parameter space of low scale seesaw mechanism. In particular, hopefully the resonant leptogenesis scenario~\cite{lowscale:LG} can be examined. 

This work is organized as follows. In Section II, we introduce the model frameworks in which the RHN pair production is important.  In Section III, three signatures of RHN pair production at the 14 TeV LHC and future 100 TeV $pp$ collider are studied. The conclusion is given in  Section IV.

\section{Simplified models with RHN pair production}



In many UV models that incorporate the type-I seesaw mechanism, RHNs participate in new interactions, Yukawa and/or gauge interactions. For our purpose, studying pairly produced RHNs at the hadronic colliders, it is enough to work in the following simplified models which can effectively describe the UV models~\cite{Kang:2015nga}:
\begin{align}
-{\cal L}_X=&\f{1}{2}M_X^2 X_\mu X^\mu+g_N\overline {N_R} \gamma_\mu   N_R X^\mu+g_q\bar q \gamma_\mu q X^\mu, \label{EFT1}\\
-{\cal L}_\phi=&\f{1}{2}m_\phi^2\phi^2+\f{\ld_N}{\sqrt{2}}\phi \overline {(N_R)^c} {N_R}+\sin \theta\f{\alpha_s}{m_t} \phi GG. \label{EFT2}
\end{align}  
We will be interested in heavy resonances, a vector boson $X_\mu$ or a scalar $\phi$, such that the pair production of RHN can be resonantly enhanced. This is well consistent with the fact that current searches on new resonance have already pushed them to the heavy region. The scalar resonance acquires coupling to gluons via its mixing with the SM Higgs boson. Viewing from the current LHC Higgs data, this mixing angle ($\theta$) is still allowed to be as large as $0.4$~\cite{Lopez-Val:2014jva}.  However, it encounters the perturbativity problem as $m_\phi$ goes into the multi-TeV region~\cite{Kang:2015nga}; we will come back to this point soon later.

As a matter of fact, the above simplified models usually are simultaneously presented in the UV completions where RHNs are charged under a new gauge group; RHNs must also couple to a scalar field which breaks the gauge symmetry to acquire Majorana masses $M_N\simeq\ld_N v_X/\sqrt{2}$, with $v_X$ the breaking scale of new gauge symmetry. It is true not only for an Abelian gauge group $U_{x B-yL}$ but also for a non-Abelian gauge group like $SU(2)_R$. Generically, the new gauge bosons (whose masses and gauge coupling strengths) are strongly restricted by the experimental data and therefore contribute to RHN pair production insignificantly. For instance, in the benchmark UV completion BLSM (see an introduction to this model in Appendix.~\ref{spectra}), $Z_{B-L}$ couples both to quarks and leptons, thus being stringently constrained by the dilepton resonance searches at the LHC~\cite{Chatrchyan:2012oaa,Aad:2014cka}. In another example, models with a local $U(1)_L$, $Z_L$ can induce the  Lagrangians in Eq.~(\ref{EFT1}) only in the case  of sizable mixing between the gauge bosons of $U(1)_L$ and $U(1)_Y$, so again we run into a similar situation as the BLSM. Therefore, in the lighter RHN region $\phi$ tends to be more important than $X$. However, in the heavier RHN region $X$ turns out to be more important.

Let us explain why the $\phi$-channel can be the dominant contribution to the RHN pair production only for a relatively light $\phi$ (thus light RHN), not significantly above the TeV scale. The arguments are from two aspects. One aspect is from perturbativity. For demonstration, we work in the models with a new local $U(1)$ which is broken by a new Higgs field $\Phi$ that develops a VEV $v_X\equiv\langle \Phi \rangle/\sqrt{2}$.~\footnote{The statement at the beginning of this paragraph is based on the assumption that Eq.~(\ref{EFT2}) is derived from UV models where RHN gains mass dynamically; relaxing it the statement may be not true.}  The spectra of this Higgs sector has been presented in Appendix.~\ref{spectra}. From Eq.~(\ref{eq:mh2}) one has $m_h^2\approx \f{\ld_1}{2} v^2-{\sin^2\theta}m_\phi^2$; see the various definitions therein. So, for a heavy $m_\phi\gtrsim 1$ TeV, keeping a sizeable mixing angle $\sin\theta\sim 0.4$ must require a large Higgs quartic coupling $\ld_1$~\cite{Kang:2012sy}; a multi-TeV $\phi$ is disfavored according to the perturbativity of $\ld_1$.

We now move to the other aspect. One can also show that, $\phi$ dominantly decaying into a pair of RHN, only happens in the relatively light $\phi$ region. A heavy $\phi$ with a relatively large mixing angle would imply a large $\ld_{12}$, which could make $\phi\ra hh$ easily dominate over other decay modes. One can estimate the condition for it not to happen. Explicitly, the decay widths of $\phi \ra hh$ and $NN$ are respectively given by 
\begin{align}\label{eq:wh}
\Gamma (\phi\ra hh)&\approx\f{\cos^6\theta}{32\pi}\L\f{\ld_{12}v_X}{4m_{\phi}}\R^2m_{\phi}~{\rm and} ~
\Gamma (\phi\ra N_RN_R)\approx \f{\cos^2\theta\ld^2_N }{32\pi}m_{\phi}\L1-\f{4M_N^2}{m_{\phi}^2}\R^{\f{3}{2}}.
\end{align} 
If aside from $\phi\ra N_RN_R$ all other decay modes of $\phi$ are inherited from the SM-like Higgs boson, then $\phi\ra W^+W^-$ is the dominant one, the partial width of which is twice $\Gamma (\phi\ra hh)$ in the high energy limit $m_{\phi}\gg m_{h}, m_W$ and the decoupling limit $\cos\theta\ra 1$. This relation is underlaid by the equivalence theorem. As a rough estimation, taking $\cos\theta\ra1$ and neglecting the phase space suppression factor, the condition for the $N_RN_R$-mode dominating over the $WW$-mode is $\ld_N\gtrsim \ld_{12}/(2\sqrt{\ld_2})$; we have approximated $m_\phi$ as $\sqrt{\ld_2/2}v_X$ in the light of Eq.~(\ref{eq:mh1}). It is illustrative to rewrite this condition as 
\begin{align}\label{}
\f{M_N}{m_\phi}\gtrsim \f{R}{2} \,|\sin\theta|=0.5\times\L\f{R}{5}\R\L\f{|\sin\theta|}{0.2}\R.
\end{align} 
Thus $R \equiv v_X/v$ can not be very large, otherwise we have $M_N>0.5 m_\phi$ for a sizable $\sin\theta$, resulting in a forbidden $\phi\ra N_RN_R$ channel. Immediately, a relatively small $R\lesssim 5$ implies a relatively light $\phi$, whose mass is $m_\phi\simeq \sqrt{\ld_2/2}R v\approx 1.7 \L \f{\ld_2}{4}\R^{1/2} \L\f{R}{5}\R$ TeV.

\begin{figure}[htb]
\includegraphics[width=1.02\textwidth]{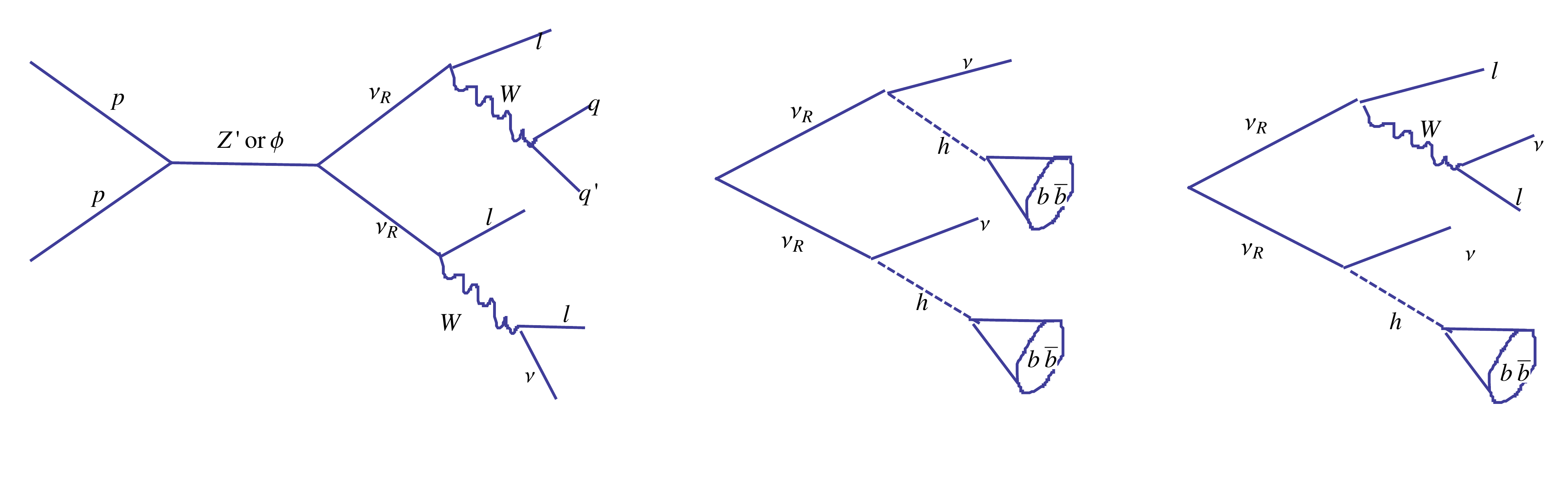}
\caption{Resonant production and decay of the RHN pair. The boosted Higgs bosons are schematically depicted within a small cone.}
\label{f:signal} 
\end{figure}

The RHN decay modes are well studied in  literatures (see for example Ref.~\cite{He:2009ua}). They can be calculated from the following Lagrangian:
\begin{align}\label{}
{\cal L}\supset& -\f{g}{\sqrt{2}} \bar\ell_{L} \gamma_\mu U_{\nu N}
N W^\mu-\f{g}{2\cos\theta_w}\bar \nu \gamma_\mu U_{\nu\nu}^\dagger U_{\nu N}NZ^\mu-
\f{h}{v} \overline{N^c} M^{diag}_{N} U^{\dagger}_{\nu N} U_{PMNS} \nu+h.c.,
\end{align} 
where the definitions of matrix $U_{\nu N}$, etc., can be found in Appendix.~\ref{mixing:U}. Here $\nu$ and $N$ denote the active and sterile neutrino in the mass eigenstates, respectively. For a TeV scale RHN, its decay well respects the equivalence theorem which leads to the following relations~\cite{He:2009ua}:
\begin{align}\label{}
&\Gamma(N_\alpha\ra  W^- \ell_i^+)=\Gamma(N_\alpha\ra  W^+ \ell_i^-)\approx \f{g^2}{64\pi M_W^2}M_N^3|(U_{\nu N})_{i\alpha}|^2.
 \cr
&\Gamma(N_\alpha\ra  Z \nu_i)\approx \Gamma(N_\alpha\ra  h \nu_i)\approx \f{g^2}{64\pi M_W^2}M_N^3|(U_{PMNS}^\dagger U_{\nu N})_{i \alpha}|^2.
\end{align} 
If the final flavors are inclusive, we can readily check that indeed the decay widths of these decay modes are equal. However, the hierarchical mixing with $|(U_{\nu N})_{3\alpha}|^2\gg |(U_{\nu N})_{1,2\alpha}|^2$ is also possible. It has important implication to the $W\ell$ mode, which then is dominated by the $\tau-$flavor and hence is not easy to be probed at LHC. Actually, the $W\ell$ mode in the $e/\mu-$flavor case has received some attention before~\cite{AguilarSaavedra:2009ik,delAguila:2009bb,Deppisch:2013cya,Abdelalim:2014cxa}. 
While the Higgs mode was just considered recently~\cite{Kang:2015nga}, focusing on the boosted Higgs region. This channel does not concern the lepton flavor, so it instead may provide the best chance in the $\tau-$flavor case. As for the $Z\nu$ mode, it is promising only for the leptonic $Z$ decay, which nevertheless is suppressed by the small branching ratio $\simeq 6.8\%$. Thus in this paper we will concentrate on three channels, di-$W$, di-Higgs and as well $h W$ to search for the RHN pair at 14 and 100 TeV hadron colliders, as shown in Fig.~\ref{f:signal}.

 \section{Collider searches}

In our collider studies, both the $X$ resonance mass $M_X$ and the RHN mass $M_{{N}}$ will be treated as free parameters. And our goal is to develop a global 
picture of the discovery prospect on the $M_{X}-M_{{N}}$ plane with 
$M_{X}>2M_{{N}}$ by studying RHN pair production with subsequent decay 
${N} \to h \nu_L$ or ${N} \to lW$. 
As the two mass parameters would pass through a wide region, the kinematic 
features of the final states will experience significant changes. 
On the other hand, it is almost  impossible to optimize cuts for each grid 
(in the $( M_{X} , M_{{N}} )$ plane) in the light of the corresponding kinematic features. Therefore, we will select five benchmark points S1-S5 (defined later), 
which are supposed to be representative for the entire patterns of kinematic 
features over the full mass region~\footnote{Of course, the cuts can be improved, even significantly, if we are restricted to a small mass region. Thus our results are fairly conservative.}. Five signal regions are obtained after optimizing the cuts 
for those benchmark points. Then, we apply the signal regions to explore 
the wide region in the $( M_{X} , M_{{N}} )$ plane
~\footnote{For each grid on the $( M_{Z'} , M_{{N}} )$ plane we will apply all of these cuts and the one that gives the best search sensitivity will be chosen.}.

The five benchmark points  as well as their basic features are presented in the 
following (hereafter we will take $X$ as $Z'$):
\begin{description}
\item[ S1: $M_{Z'} = 0.35$ TeV \& $M_{{N}} =$ 150 GeV]  Both particles are light, so neither ${N}$
nor its secondary decay product $h/W$ is boosted. The resulting jets and leptons are relatively soft, and thus {are easy to be buried in the SM backgrounds}.
\item[ S2: $M_{Z'} = 1.45$ TeV \& $M_{{N}} = $700 GeV] Both are heavy but $Z'$ mass is near the threshold  $2M_{{N}} $ such that ${N}$ is non-boosted; $h/W$ is well boosted due to the heaviness of RHN, so the angular separation between the decay products of $h/W$ typically are small:
\begin{align}
\Delta R_{b\bar{b}({jj})} \sim \frac{2 m_{h/W}}{p_T(h/W)} \sim \f{m_{h/W}}{\frac{M_{{N}}}{2}\L1-{m_{h/W}^2}/{M_{{N}}^2}\R} \sim 0.7(0.5).
\end{align}
This feature enables us to tag the Higgs jet using the jet substructure technique~\cite{Butterworth:2008iy}.  

\item[ S3: $M_{Z'} = 1 $ TeV \& $M_{{N}} = 150$ GeV]  {For this pattern, the produced RHNs are  highly boosted and travel back to back. On the other hand, since the mass difference between ${N}$ and its decay products $h/W$ is small, it rises problems both in the ${N} \to h \nu_L$  and ${N} \to l W$ channels. For the former, two neutrinos just like the RHN pair are flying back to back, thus rendering a small vectorial \ET. As for the latter channel, the angular separation between the lepton and $W$ is fairly small,
\begin{align}
\Delta R_{lW} \sim \frac{2(M_{{N}} - M_W)}{p_T({N})} \sim2\times \f{M_{{N}} - M_W}{\sqrt{\frac{M_{Z'}^2 - 4 M_{{N}}^2}{4}}} \sim 0.3.
\end{align}
As a consequence, the lepton tends to be non-isolated because it is too close to either another lepton from a leptonically decaying $W$ ($W_\ell$) or the jets from a hadronically decaying $W$ ($W_h$).  
 }

\item[  S4: $M_{Z'} = 5$ TeV and $M_{{N}} = 700 $ GeV]  Both ${N}$ and $h/W$ from ${N}$ decays are  boosted. Their heaviness make the jets/leptons in the final states energetic. Moreover, despite of the boosted RHNs, the hard lepton from ${N}\ra \ell W$ does not suffer a serious isolation problem, because their angular separation is $\Delta R_{lW} \sim0.5$.

\item[ S5: $M_{Z'} = 10.05$ TeV and $M_{{N}} = 5$ TeV]  We design this benchmark point to represent super boosted $h/W$, whose cone size is estimated to be
\begin{align}
\Delta R_{h,W} = \frac{2 m_{h,W}}{p_T(h,W)} \sim \mathcal{O}(0.1),
\end{align}
which is even smaller than current jet area resolution at LHC, $\sigma(R) \sim 0.2$~\cite{Spannowsky:2015eba,Bressler:2015uma}. In this case, neither $h$ nor $W$ shows any substructure and we can only observe a narrow jet with a relatively large invariant mass. 

\end{description}

We adopt the UFO model files of $U(1)$ extended SM~\cite{Basso:2011na}, 
written by FeynRules~\cite{Alloul:2013bka}. The monte carlo events for the signal 
and backgrounds are generated through MadGraph5\_aMC@NLO~\cite{Alwall:2011uj}, in which the Pythia6~\cite{Sjostrand:2006za} is used for decaying SM particles, parton showering and hadronization. The Delphes3~\cite{deFavereau:2013fsa} with default ATLAS setup is chosen for our fast detector simulation. We adopt the same $b$-tagging method as in Delphes throughout the analysis in 
this paper. Concretely speaking, a jet is tagged as a $b$-jet with probability 
70\% if a parton level $b$ quark is found within the cone with size $\Delta R =0.3$ 
centred on the jet direction; otherwise, the $b$-tagging probability is 
20\% or 0.5\% depending on if a charm quark is found or not.

The analysis procedure on each benchmark point is as the following. First, we apply some preselection cuts to guarantee the existence of certain objects, which are necessary for the reconstruction of kinematic variables. Then, we feed back these variables to the TMVA package in the ROOT and calculate the multi-variable analysis (MVA) response distribution. Concretely, the BDT method is used for MVA. Finally, we impose a cut on the BDT variable such that the signal significance is maximized 
and at the same time sufficient signal events are retained. In the rest of this section 
we will first study the di-$W$, di-Higgs and $hW$ channels case by case at the 14 TeV LHC and the 100 TeV future $pp$ collider, and at last combine them to find the farthest reach on RHN search in the BLSM.

\subsection{The di-$W$ channel: trileptons}

For this channel, both RHNs decay into $\ell W$ resulting in the final states $W^\mp W^\mp \ell^\pm \ell^\pm$. It is the most promising channel and presents three remarkable signatures, same-sign dilepton (SSDL), trilepton~\footnote{This signature was studied before the LHC era, merely restricted to a few benchmark points~\cite{Basso:2008iv,delAguila:2009bb,Perez:2009mu}. Our work is not only a timely revising at the LHC era but also a big update towards the future.} and four leptons. Note the SSDL mode only occurs for Majorana RHN, while the trilepton and four lepton modes occur for both Dirac and Majorana RHN. The individual branching ratios for Majorana RHN can be estimated as follows: 1) For SSDL, only half of the final states contribute to it and moreover both $W$ bosons should decay hadronically, thus giving rise to a suppression factor $\frac{1}{2} \times {\rm Br}^2(W \to jj)$; 2) for trilepton, any combinations of final states are allowed, but one $W$ should decay leptonically, leading to a suppression $2 \times {\rm Br}(W \to jj) \cdot {\rm Br}(W \to \ell \nu)$; 3) for the four lepton, obviously it has a suppression ${\rm Br}^2(W \to \ell \nu)$. Therefore, we obtain their relative ratios
 \begin{align}
{\rm Br}(3\ell): {\rm Br}(\ell^\pm\ell^\pm): {\rm Br}(4\ell) \sim 12:9:2,
\end{align}
where $\ell = e, \mu$ and we have used ${\rm Br}(W \to \ell \nu) \sim 2/9$ and ${\rm Br}(W \to jj) \sim 6/9$.
Moreover, for SSDL, the non-prompt leptons from $t\bar t$ provides a robust BG~\cite{AguilarSaavedra:2009ik,delAguila:2009bb,kang:2014jia}; for $4\ell$, the existence of two neutrinos and ambiguity of combining the four leptons render the mass reconstruction of RHN impossible. Therefore, we will study the trilepton signature in this work.~\footnote{Recently, Ref.~\cite{Abdelalim:2014cxa} estimated the search sensitivities for the 2$\ell$ +jets, 3$\ell$ +jets and 4$\ell$ signatures, respectively, considering no detector effects. They found that the 3$\ell$ +jets signature is the worst. However, it may be not true after considering the finite detector resolution; for example, $ZZ$+jets and the irreducible BGs like $ttV$ will become quite significant BGs for 4$\ell$.}

\subsubsection{Backgrounds and pre-selection}\label{prese}

For this trilepton plus jets signature, the NLO production cross sections of its main BGs at the 
14 (100) TeV proton-proton collision are listed in the the 
 second  column of Table~\ref{preww}. 
The di-boson BGs are generated with up to two additional jets at parton level since we require 
at least two jets in the final state. To avoid double counting between the matrix element 
calculation and the parton showering, we turn to the MLM matching method, taking an appropriate xqcut for each di-boson BG. 
Those BGs involving a leptonic $Z_\ell$, especially $W_\ell Z_\ell$, 
constitute the dominant BGs, because the requirement on lepton number is easy to fulfill there. 
For the $V_\ell Z_\ell$-BG, the jets are mainly from the initial state radiation (ISR). 

We apply the following pre-selection cuts: A)  at least three leptons; B) at least two (one) jets 
at 14 (100) TeV; C) no $b$-tagged jets, which is useful to suppress the large BGs that contain 
top quarks. Note that in order to keep the signal events with two collinear jets from the  highly boosted $W$ decay, we only require one jet at 100 TeV; such a treatment is particularly 
important with respect to S5. After the preselection, the main BGs are $W_{\ell} Z_{\ell}$, 
$Z_{\ell} Z_{\ell}$ and $t\bar{t}Z$; see the second column of Table~\ref{preww}. 
It is seen that from 14 TeV to 100 TeV, all the cross sections of BGs (after pre-selections), 
in particular the dominant BG $V_\ell Z_\ell$, increase  by (more than) an order of magnitude. 
One of the main causes is that  one less jet is required. 
 \begin{table}[htp]
  \begin{tabular}{|c|c|c|} \hline
    & $\sigma_0 (\rm pb)$ & ~~~ $\sigma_{pre} (\rm fb)$~~~~\\ \hline
  $t\bar{t}W$ & 0.67[16.6] & $0.52[3.3]$  \\
  $t\bar{t}Z$ & 0.93[55] & 2.2[39.6] \\
  $W_\ell  W_\ell W_\ell $ &$1.3 \times10^{-3}[8.7\times10^{-3}] $ & 0.05[1.7]\\
  $W_\ell  Z_\ell $ & 0.36[2.44] & 8.2[359.8]\\
  $Z_\ell  Z_\ell $ & 0.029[0.17] & 2.1[61.8] \\
  $HW$ & 1.57[14.8] & 0.09[2.5]\\
  $t\bar{t}H$ & 0.56[31.6] & 0.27[4.3] \\ \hline
  S1 &1  & 59.2[45.2] \\ \hline
  S2  & 1 & 132[106.5] \\ \hline
  S3  & 1& 44.6[35.0]\\ \hline
  S4 & 1 & 140[178.1] \\ \hline
 S5 & 1 & NO[212.9] \\ \hline
    \end{tabular}~~~~~~~
       \caption{\label{preww} Di-$W$ channel: Cross sections for signals and backgrounds before (second column) and after preselection (third column). All the signal production cross sections have been normalized to 1 pb. As a convention used throughout this paper, the quantities outside and inside the square brackets are for the 14 TeV and 100 TeV cases, respectively. ``NO" means that there are no corresponding quantities at 14 TeV.}
\end{table}

\begin{figure}[htb]
\begin{center}
\includegraphics[width=0.47\textwidth]{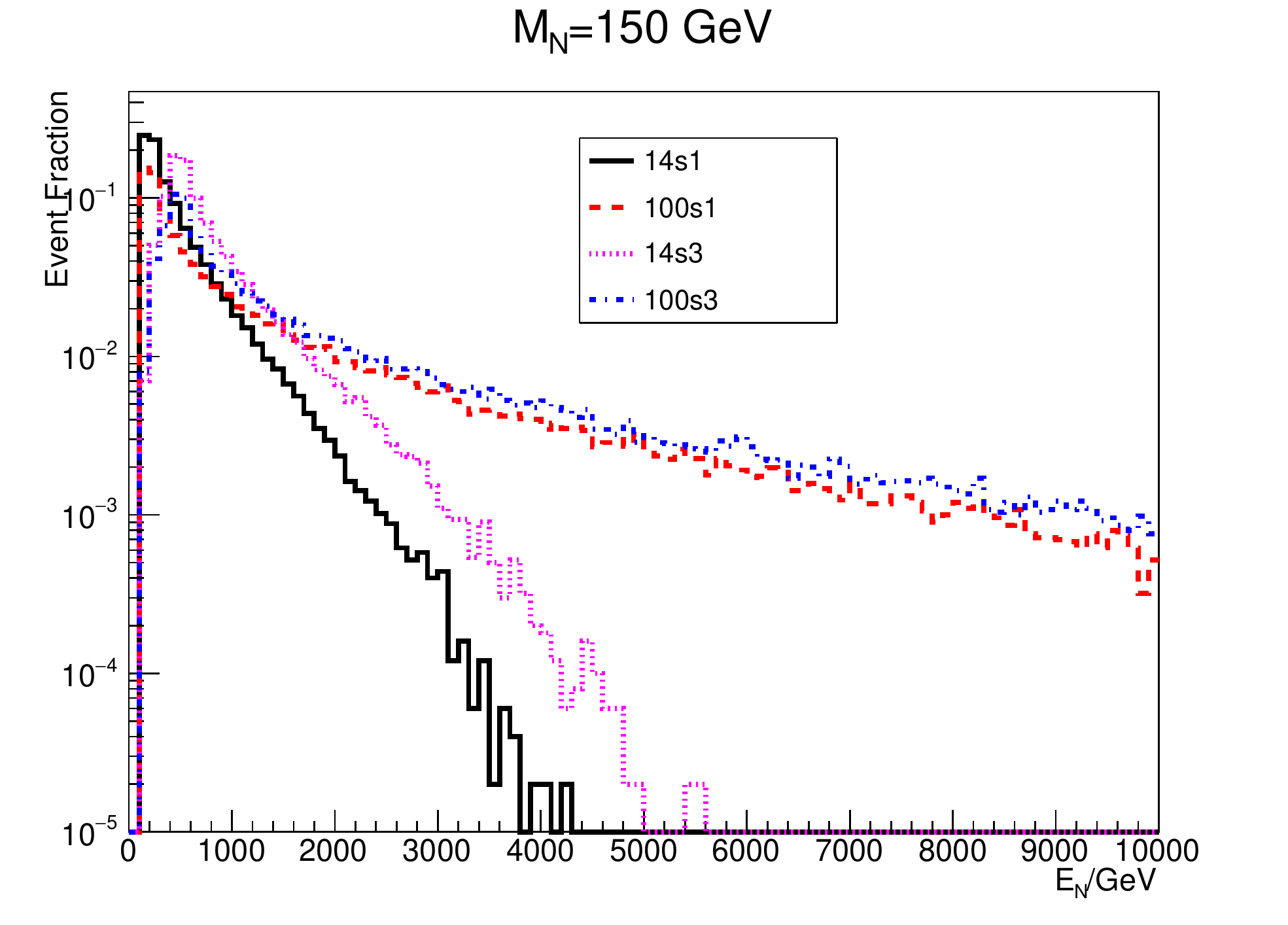} 
\includegraphics[width=0.47\textwidth]{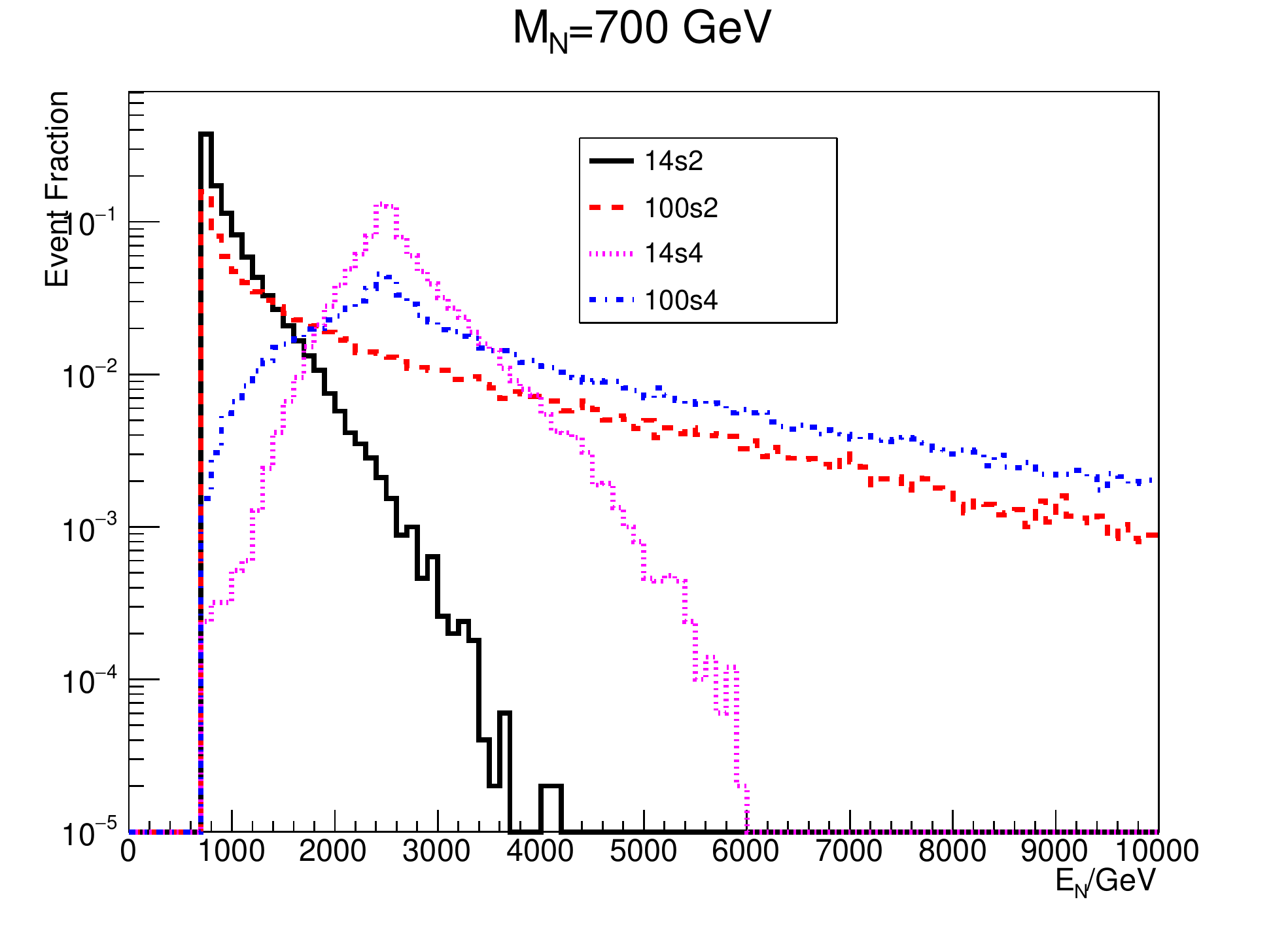} 
\end{center}
\caption{\label{fig:vre} The distributions of the RHN energy $E_{N}$, at 14 and 100 TeV for S1-S4. }
\end{figure}
The cut efficiencies for 4/5 benchmark points 
are also given in Table~\ref{preww}, where we have assumed a common nominal production 
cross section of 1 pb for all benchmark points in $pp$ collisions at 14 and 100 TeV. 
As we have expected, the cut efficiencies are relatively low for S1 and S3: for S1, some of the final states are too soft to be reconstructed; as for S3, the primary lepton from RHN decay suffers from the isolation problem. The preselection reduces the number of signal events 
by around one order of magnitude for S2 and S4. The signal preselection efficiencies tend to decrease from 14 to 100 TeV,  since the more energetic RHN (one can see it from Fig.~\ref{fig:vre}), renders the final leptons difficult to be isolated. One can also see this from the cut flow in Table~\ref{vre}: at 100 TeV the $n_\ell$ cut becomes more stringent for all benchmark points due to the severer lepton isolation issue.
 \begin{table}[htb]
  \begin{tabular}{|c|c|c|c|c|c|c|c|} \hline
   14[100] TeV(/50000)  & s1 & s2 & s3 & s4\\ \hline
   $n_\ell >2[2]$ & 4356[2734] & 9829[5989] & 2645[2000] & 13267[9723] \\
   $n_j>1[0]$  & 2216[2382] & 6543[5706] & 2001[1837] & 7811[9423] \\
   $n_b=0[0]$ & 2123[2307] & 6065[5366] & 1899[1749] & 7286[8697] \\ \hline 
     \end{tabular}~~~~~~~
       \caption{\label{vre} Preselection cut flow for benchmark points with 50000 total number of events.}
\end{table}

\subsubsection{Multivariable analysis}

After selecting the events with required objects (at least 3 lepton and 2 jets at the 14 TeV LHC), we are able to reconstruct kinematic variables which are used to discriminate between signal and BGs. In order to obtain the best discrimination power through quite a few correlated variables, we will employ MVA. In this channel, the variables adopted in MVA are 
\begin{align}\label{WWpre:14}
& N_\ell, ~ N_j,~m_{j_1,j_2}, ~p_T(\ell_1), ~p_T(j_1),\nonumber \\
& p_T(\ell\ell), ~p_T(\ell jj), ~\eta(\ell\ell), ~\eta(\ell jj), ~\phi(\ell\ell), ~\phi(\ell jj), ~m_T(\ell\ell), ~m_T(\ell jj),\nonumber \\
& m_{\ell\ell}, ~m_{\ell jj}, ~m_{all}, ~m_{T_2}(\ell\ell,\ell jj), 
\end{align}
where $N_{\ell/j}$ is the number of leptons/jets with jets reconstructed using the anti-kt algorithm~\cite{Cacciari:2008gp}
with $R=0.4$. $m_{\ell \ell}$ is the invariant mass for the dilepton system and $m_T(\ell \ell) = 2 E_T(\ell_1) E_T(\ell_2) (1- \cos \phi(\ell_1,\ell_2))$ is the transverse mass of it; here $m_{T_2}$ is defined as~\cite{Lester:1999tx,Barr:2003rg}
\begin{align}
m_{T_2}(\ell\ell,\ell jj)=  \min_{\slashed{p}_T^1 + \slashed{p}_T^2 = \ET } \ [ ~ \max(m_T(\ell\ell, \slashed{p}_T^1),m_T(\ell jj, \slashed{p}_T^2) ) \ ],
\end{align} 
which shows a kinematic edge at $M_{{N}}$ and thus is quite helpful for a heavy RHN; finally, $m_{\ell jj}$ is the reconstructed RHN mass using the hadronically decaying $W$.

In this trilepton signature, two of the three leptons along with \ET are from one RHN decay while the third lepton along with two leading jets are from the other RHN decay. Reconstructing these two sub-systems, the di-lepton and di-jet subsystem, not only helps much to overcome BGs but also enables us to estimate the RHN mass. However, the way of combining among the three leptons is not unambiguous and we propose three methods in the following:
\begin{itemize}
\item The closest two leptons on the $\eta-\phi$ plane are identified as the di-lepton subsystem; the third lepton is combined with two jets to form the di-jet subsystem.
\item The lepton closest to the di-jet subsystem on the $\eta-\phi$ plane is combined with two jets; the rest two leptons form the di-lepton subsystem. 
\item Figure out the combination that gives the longest angle distance between two subsystem, 
${\rm Max}\{\sum_{ijk}\Delta R(\ell_i\ell_j,\ell_k jj)\}$ with $i$, $j$ and $k$ different than each other.
\end{itemize}
In the practical operation, we shall try all the methods and the one that gives the largest signal significance after MVA analysis will be selected out. It is found that the third method stands out in most cases.


We insert a discussion on the situation at 100 TeV. The pre-selection cuts are the same as the 14 TeV case except that we require only one jet instead of two, on account of the super boosted hadronic $W$ actually behaving as a single jet. This time the variables that we used for MVA are chosen as the following 
\begin{align}\label{WWpre:100}
&n_\ell, ~n_j, ~m_{j_1,j_2}, ~p_T(\ell_1), ~p_T(j_1), ~p_T(\ell\ell), ~p_T(\ell jj),  ~m_{\ell\ell}, ~m_{\ell jj},  \nonumber \\ 
&m_{\ell\ell \ell jj}, ~m_{T_2}(\ell\ell,\ell jj),  \nonumber \\
&p_T(\ell_2), ~p_T(\ell_3), ~p_T(j_2), ~m_{j_1},  p_T(\ell\ell), ~p_T(\ell j),  ~m_{\ell\ell}, ~m_{\ell j}, ~m_{\ell\ell\ell j}, ~m_{T_2}(\ell\ell,\ell j).
\end{align}
The variables in the last line are specific to the 100 TeV collider, and they are constructed in the presence of only one jet;~\footnote{Note here that the method used to identify the di-lepton and 
di-jet subsystems requires a slight modification on the previous one, considering actually only 
one jet is present.} other variables are similar to the 14 TeV case. 

Now we feed back all variables listed in Eq.~(\ref{WWpre:14}) or Eq.~(\ref{WWpre:100}) to TMVA. The BDT method is used to train these discriminators. To have a better understanding of the role that each kinematic variable plays in BDT, we list the five most important variables for each signal region; see Table~\ref{Top:wl}. We have quite a few remarks in orders:
\begin{table}[htp]
\begin{tabular}{|c|c|c|c|c|}\hline 
   S1 & S2 & S3 & S4 &S5\\ \hline
   $m_{\ell\ell \ell jj}[m_{\ell \ell }]$ & $p_T(\ell_1)[p_T(\ell_1)]$ & $p_T(\ell jj)[m_{\ell \ell \ell j}]$ & $m_{\ell \ell \ell jj}[p_T(\ell j)]$ & NO$[m_{\ell \ell \ell j}]$ \\
   $m_{\ell \ell } [p_T(j_1)]$ & $m_{\ell \ell \ell jj}[m_{\ell \ell \ell j}]$ & $m_{\ell \ell \ell jj}[m_{\ell \ell}]$ & $p_T(\ell_1)[p_T(\ell_1)]$ & NO$[p_T(\ell_1)]$ \\
   $p_T(\ell_1)[m_{\ell \ell \ell j}]$ & $m_{\ell \ell }[m_{\ell \ell }]$ & $m_{\ell \ell }[p_T(\ell j)]$ & $p_T(\ell jj)[p_T(\ell\ell)]$ &NO$[p_T(\ell j)]$ \\
   $p_T(j_1)[n_j]$ & $p_T(\ell jj)[p_T(\ell_2)]$ & $p_T(\ell_1)[p_T(j_1)]$ & $m_{\ell \ell }[m_{\ell \ell \ell j}]$ &NO$[p_T(\ell\ell)]$ \\
   $m_{jj}[m_{T_2}(\ell\ell,\ell j)]$ & $p_T(\ell \ell )[p_T(\ell_3)]$ & $p_T(j_1)(p_T(\ell_1))$ & $m_T(\ell jj)(p_T(\ell jj))$ &NO$[m_{T_2}({\ell \ell},{ \ell j})]$\\ \hline 
  \end{tabular}
\caption{\label{Top:wl}  Di-$W$ channel:  top-5 variables in BDT anslysis. }
\end{table}
\begin{itemize}

\item  The invariant masses of the total visible objects $m_{all}$ and the di-lepton subsystem $m_{\ell \ell}$, which respectively reflect the mass scale of $M_{Z'}$ and $M_{{N}}$, always are powerful discriminators. 

\item Although $m_{\ell jj}$ gives a more exact mass of RHN, $m_{\ell j}$ usually takes 
the higher rank. The reason is twofold. Firstly, compared to leptons, the worse energy resolution 
for (especially less energetic) jets leads to a relatively widespread distribution for $m_{\ell jj}$. 
Secondly, in reconstructing the $\ell jj$ subsystem, the jets, which are supposed to be from $W$ decay, 
may be hard ISR jets, thus giving a wrong $\ell jj$ subsystem. 

\item  The transverse momenta of the leading jet/lepton or the subsystems also play important roles. This is well expected, because they show main features of the heavy spectrum. 

\item The variable $m_{\ell \ell \ell j}$ is specified to 100 TeV and designed to capture the 
 highly-boosted $W$ from S5, but it is also a better discriminator than $m_{\ell \ell \ell jj}$ in other signal regions. 
This is because in constructing $m_{\ell \ell \ell jj}$, even though the leading jet tends to originate from the $W$ decay, the second leading jet is usually from ISR. Consequently, the variable using one less jet turns out to be better. Similarly, $m_{T_2}(\ell \ell,\ell j)$ is a better discriminator than $m_{T_2}(\ell \ell,\ell jj)$. 

\end{itemize}
 
 After training the discriminators with BDT method, we apply cuts on the BDT response 
 for the signal and BGs, and  the results are shown in Table~\ref{c2res}. In this table, the signal cut efficiencies 
 $\epsilon_s$ (SIG) and the BG cross sections after the BDT cuts, $\sigma({\rm BG})$, 
 are also listed. In the last column, we give the signal reaches at the 3000 fb$^{-1}$, 
whcih is  defined as :
\begin{align}
\text{signal reach} = 3.0 \times \sqrt{B+(0.05 B)^2}/(\mathcal{L}\epsilon_s),
\end{align}
where $\mathcal{L}$ is the luminosity (3000$^{-1}$ fb is used throughout the work) and $B=\mathcal{L} \times\sigma({\rm BG})$ is the total number of background events. In other words, it is the cross section required for discovery  at 3$\sigma$ level. At the 14 TeV LHC, the signal reach limits are ranked as $S4<S2<S3<S1$, which corresponds well with our expectation. The main explanations are already addressed before, in particular in Section~\ref{prese} where we understand the results after preselection. 
 \begin{table}[htp]
 \begin{tabular}{|c|c|c|c|c|c|c|c|c|c|c|c}\hline 
 \quad\quad\quad\quad  &  Cut & $\epsilon_s$(SIG) & $\sigma$(BG)(fb) & signal reach(fb)@ 3000 fb$^{-1}$ \\ \hline
S1 & BDT$>$0.2[0.2] & 0.0195[0.013] & 0.27[9.9]& 2.5[115] \\ \hline
S2 & BDT$>$ 0.5[0.4] & 0.069[0.06] & $7.9 \times 10^{-3}$[0.52] & 0.073[1.5]\\ \hline
S3 & BDT$>$0.4[0.3]& 0.016[0.016]& $5.0 \times 10^{-3}$[0.96]& 0.25[9.6]\\ \hline
S4 & BDT$>$0.6[0.5]& 0.125[0.15]&  $8.8 \times 10^{-4}$[0.062]& 0.013[0.11]\\ \hline 
S5 & BDT$>$NO[0.5]& NO[0.21]&  NO[0.042]& NO[0.061]\\ \hline 
  \end{tabular}
\caption{\label{c2res} Di-$W$ channel: cuts efficiencies and signal reaches. }
\end{table}

We can also see from the table that as the colliding energy jumps from 14 TeV to 100 TeV, the signal reaches are increased for all benchmark points (namely worse search sensitivities). The reasons, as pointed out earlier, are due to the much larger BG cross sections and more serious collimation problem of the decay products of RHN. Additionally, we can see that the relative orders of signal reaches do not change, except that now S5 has the best (moderately better than S4) search sensitivity -- its highly boosted final states can be easily distinguished from the background events.~\footnote{ Of course, in a concrete model,  points like S5 will be quite difficult to probe since they usually have very small production rates.}

\subsubsection{Digress on the PDF effects}

As stated before, some features of our results can be traced back to the PDF effect, 
so we briefly introduce it here. 
The master formula for the RHN pair production at the proton-proton collider is
\begin{align}
\sigma = \sum_{i,j} \int^1_0 dx_1 dx_2 f_i(x_1,\mu) f_j({ x_2},\mu) \hat{\sigma}_{ij\to {N} {N}},
\end{align}
where $\hat{\sigma}_{ij}$ is the parton level scattering cross section and the parton distribution function $f_i(x,\mu)$ gives the probability of the  parton $i$ that have energy fraction $x$ inside proton at energy scale $\mu$. 
For $\mu \sim \mathcal{O}(1)$ TeV, the valence quark PDF start to drop dramatically when $x \gtrsim 0.2$.

\begin{figure}[htb]
\begin{center}
\includegraphics[width=0.49\textwidth]{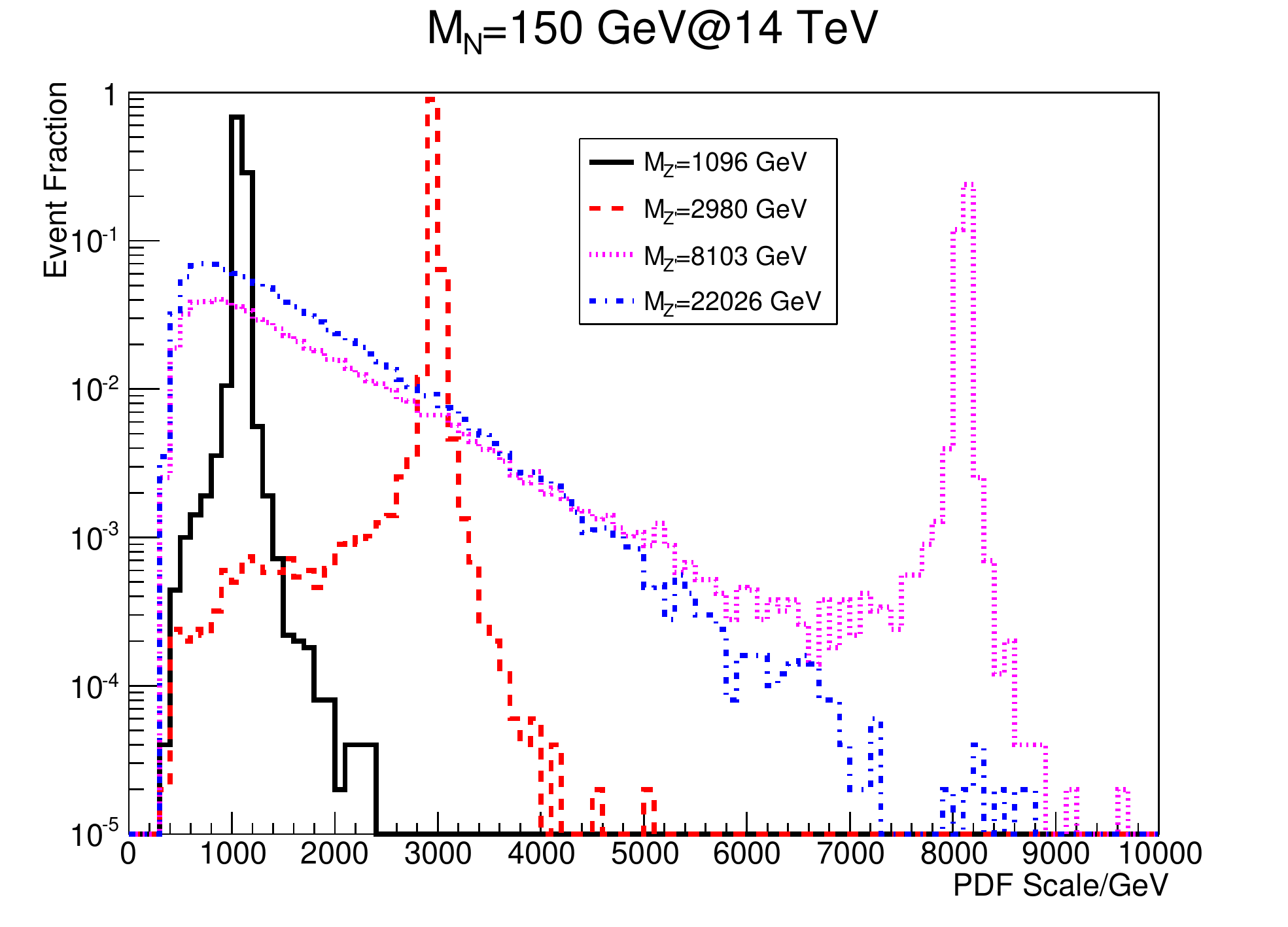}
  \includegraphics[width=0.49\textwidth]{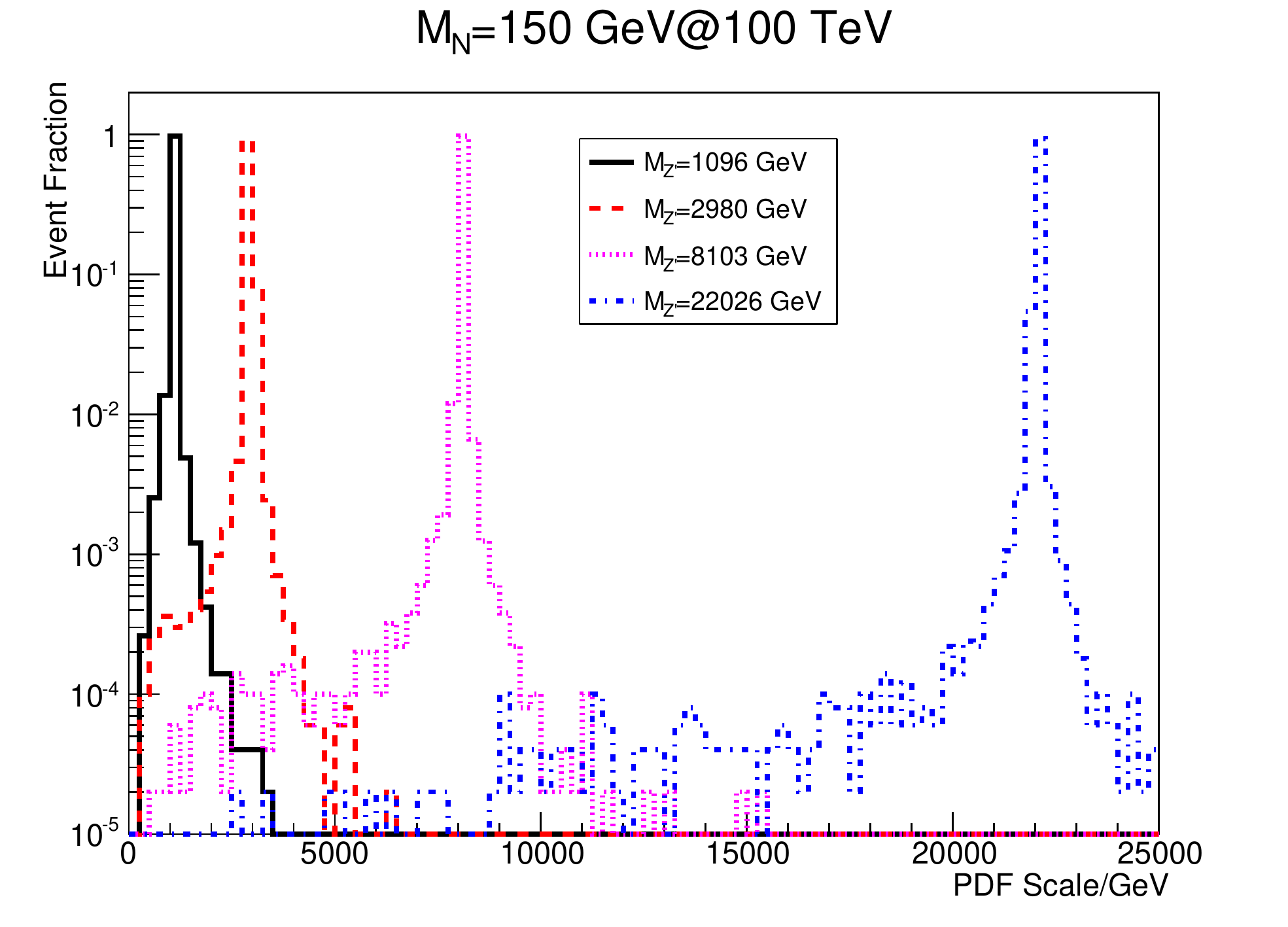} \\
    \includegraphics[width=0.49\textwidth]{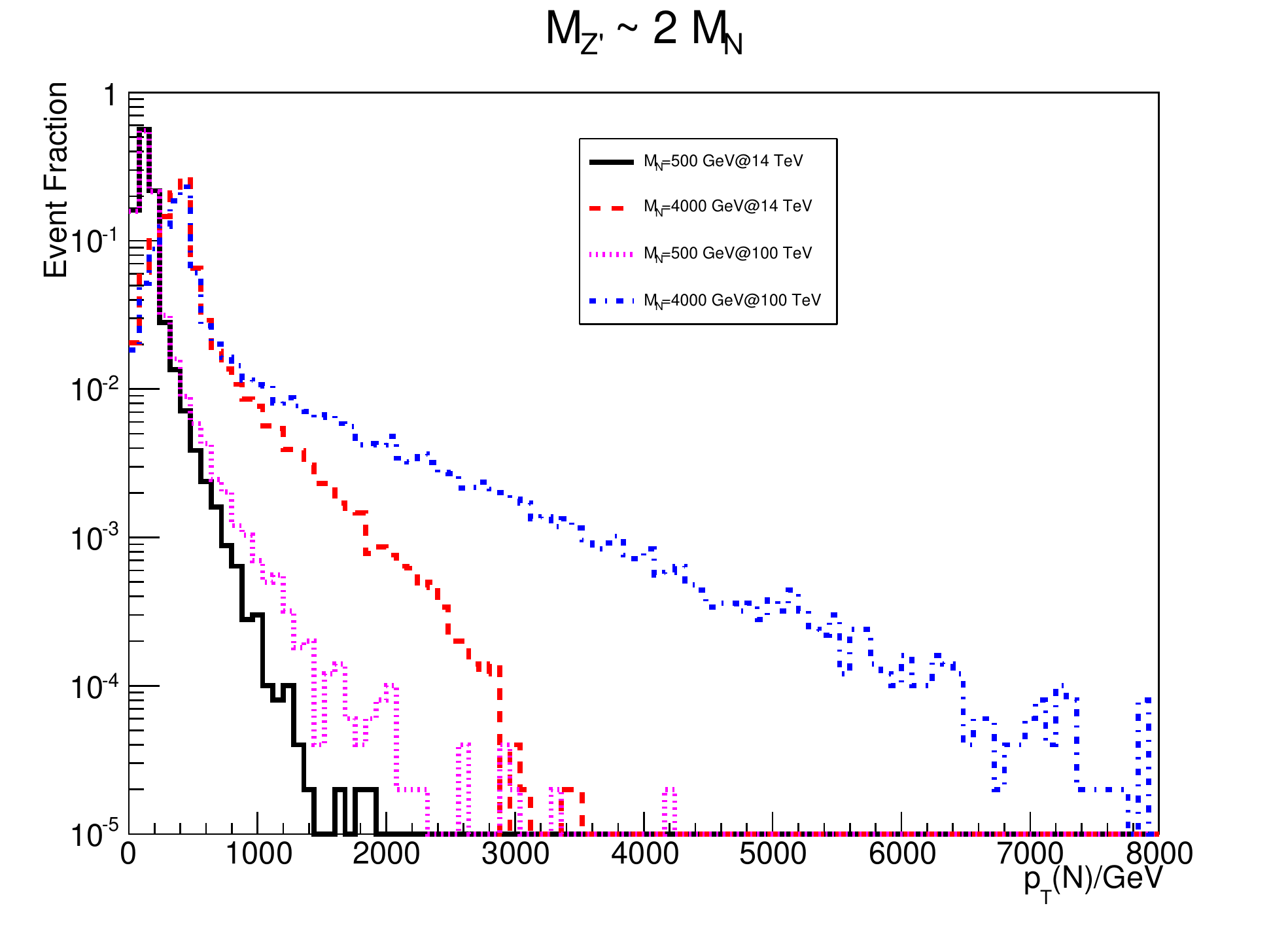}
    \includegraphics[width=0.49\textwidth]{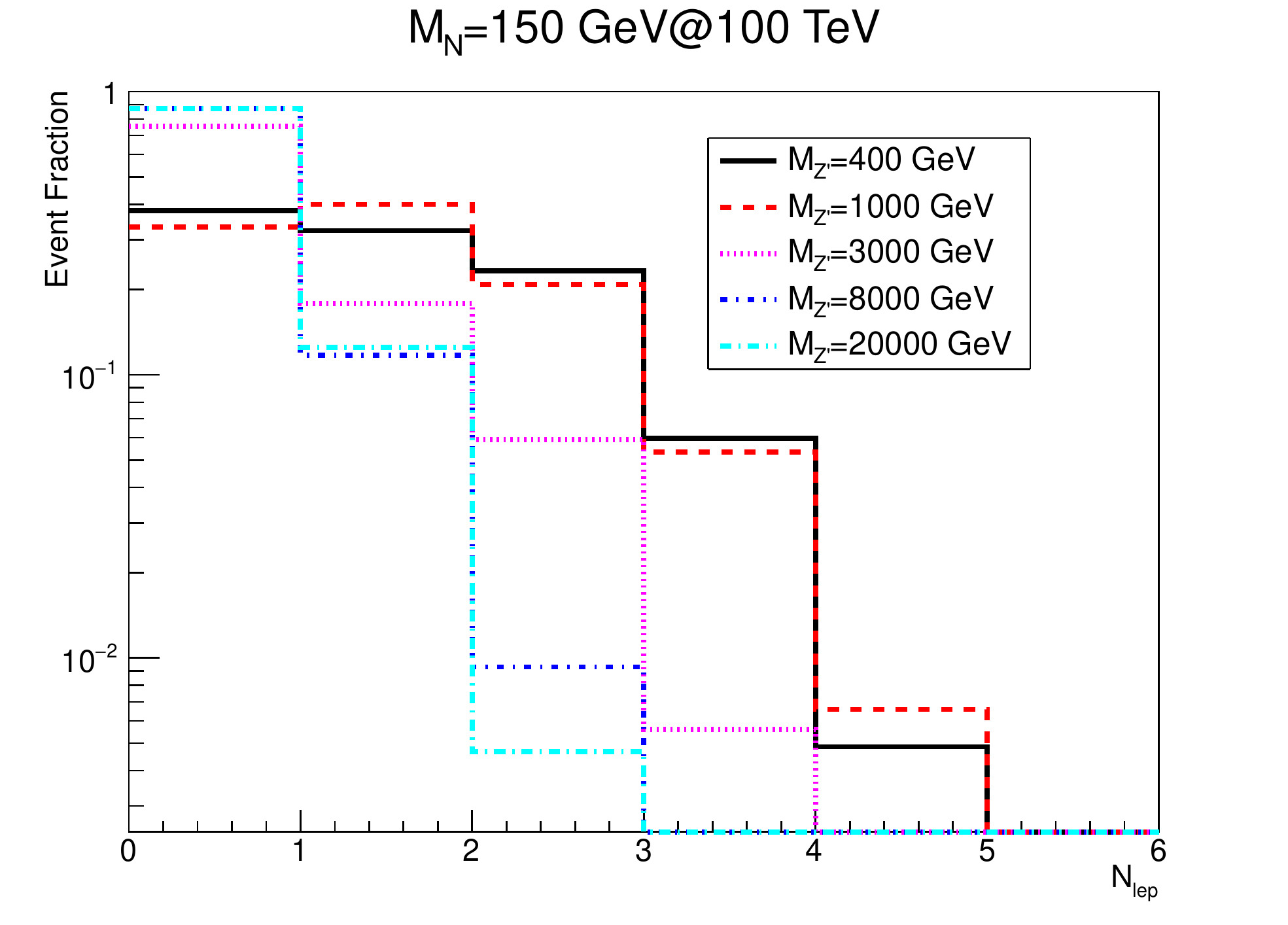} 
\end{center}
\caption{Top panels show the PDF scale distributions for different $Z'$ mass, at 14 TeV and 100 TeV.  Lower left: RHN transverse momentum distribution with given $M_{Z'} \sim 2 M_N$. Lower right: Lepton number distribution for different RHN boosts. 
}
\label{PDF} 
\end{figure}

The PDF effect becomes important for a quite heavy resonance beyond the typical CM energy of hadron collider and then partons with small $x$ tend to dominate over the production of RHN via the off-shell rather than the on-shell resonance. To have a closer look at this, we show the distributions of PDF scales (namely the total energy of the RHN pair) for the production of a light RHN pair (150 GeV) with different $Z'$ masses at 14 TeV and 100 TeV in the upper left and upper right panels of Fig.~\ref{PDF}, respectively. Since our process is dominated by the valence quark scattering at 14 TeV, the PDF effect starts to show up from $\sim 0.2\times14=2.8$ TeV.  
The distribution for $M_{Z'}=2980$ GeV (red) only shows a small bump at $\sim$ 1TeV. However, the distribution for $M_{Z'} \sim 8$ TeV (pink) has a significant jump at the low energy ($\gtrsim 2M_N$), because here the off-shell contribution to RHN pair production is comparable with the on-shell contribution.
$M_{Z'} = 22$ TeV is a limiting case with $Z'$ inaccessible at the collider thus the effective operator $\f{1}{\Ld^2}(q\bar\Gamma q)(\bar {N} \Gamma {N})$ being good enough to describe the model; RHN production is indeed dominated by partons with low $x$ (see the blue curve). Moving to 100 TeV, the PDF effect becomes very small through out the full region of our interest; see the top right panel. As a comparison, in the lower left panel of Fig.~\ref{PDF} we show the case with a heavy RHN, e.g., close to $M_{Z'}/2$, we can see that the peaks of $p_T$(RHN) distributions always follow the resonance.

\subsubsection{Results and analysis}

\begin{figure}[htb]
\begin{center}
\includegraphics[width=0.49\textwidth]{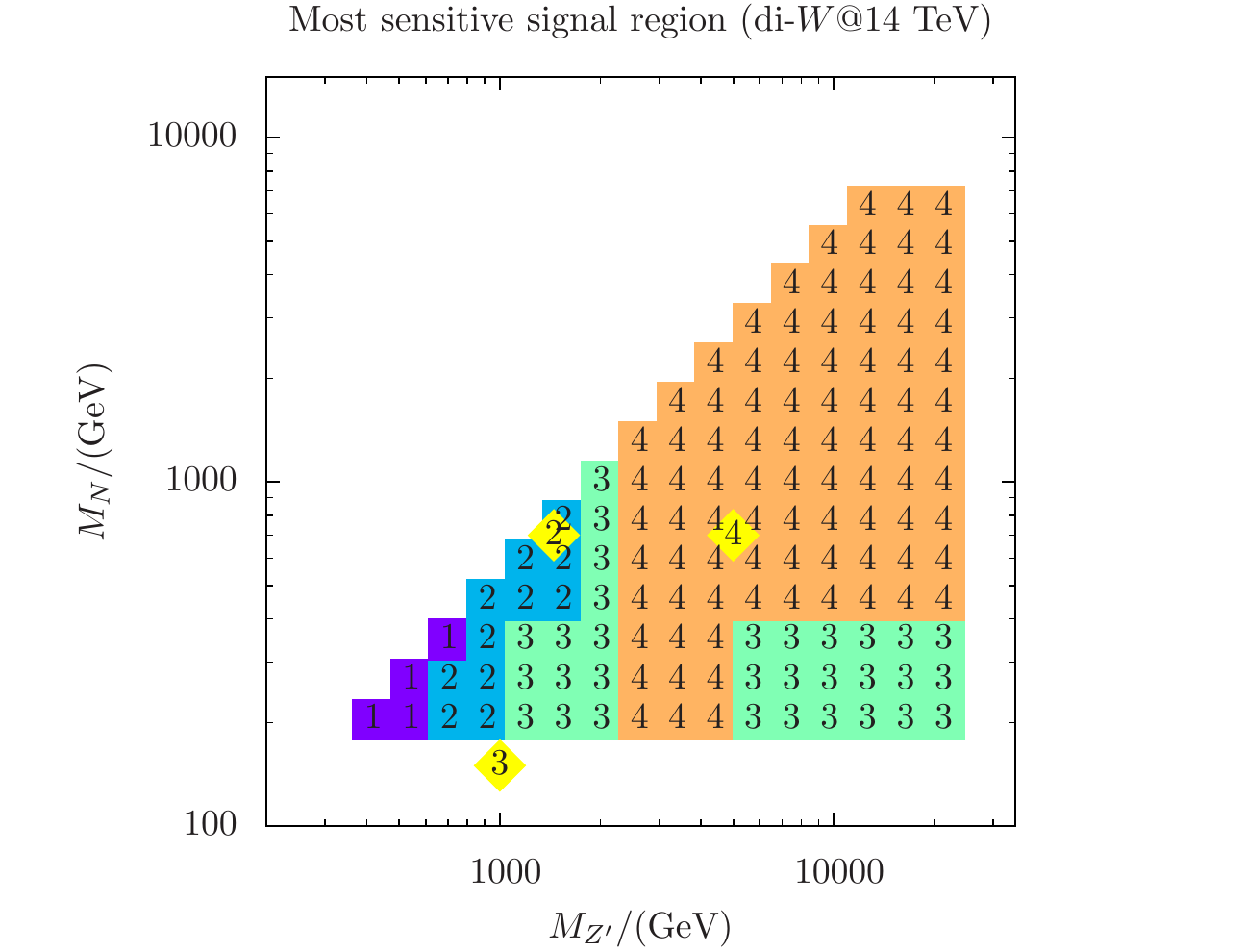}
\includegraphics[width=0.49\textwidth]{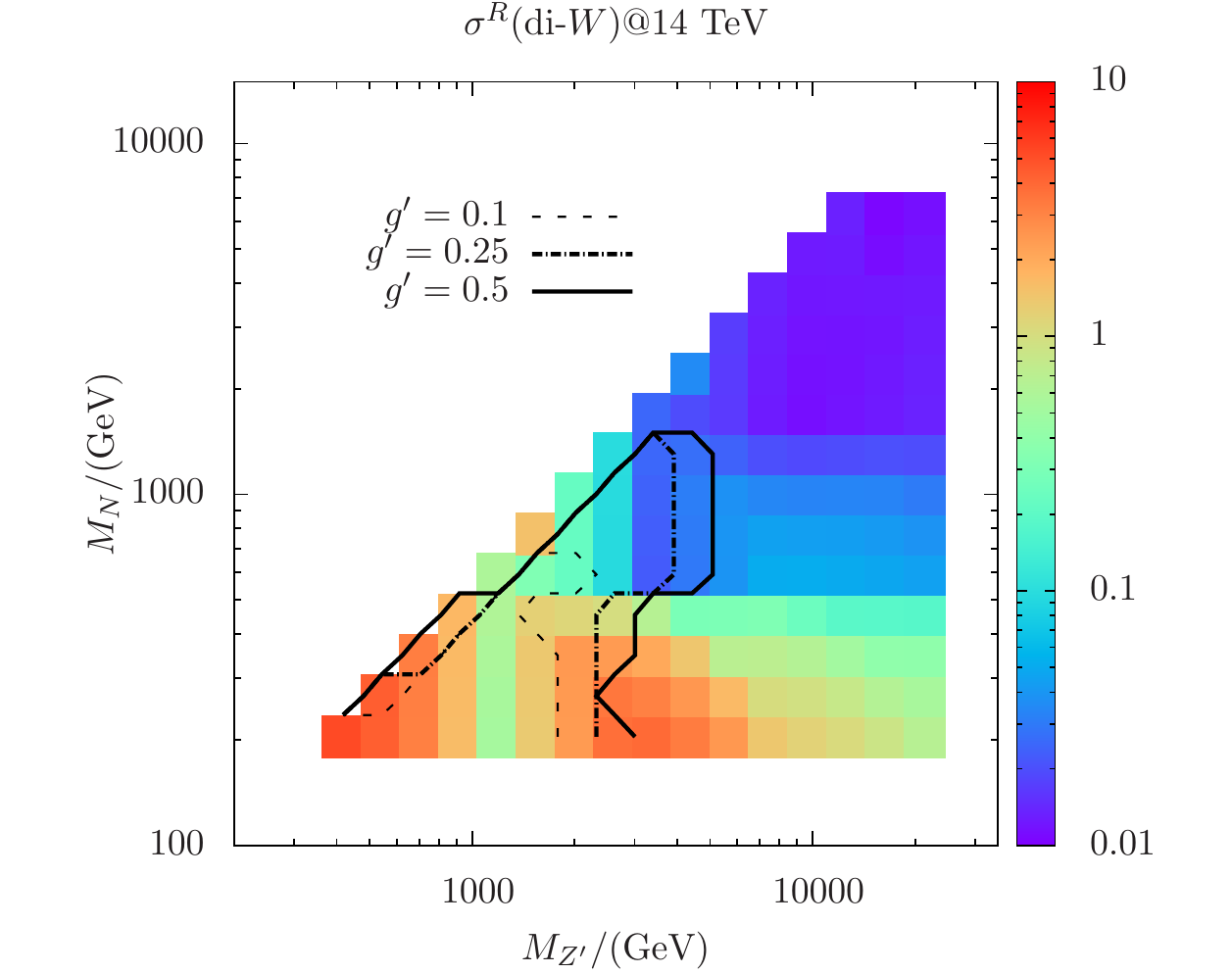} 
\includegraphics[width=0.49\textwidth]{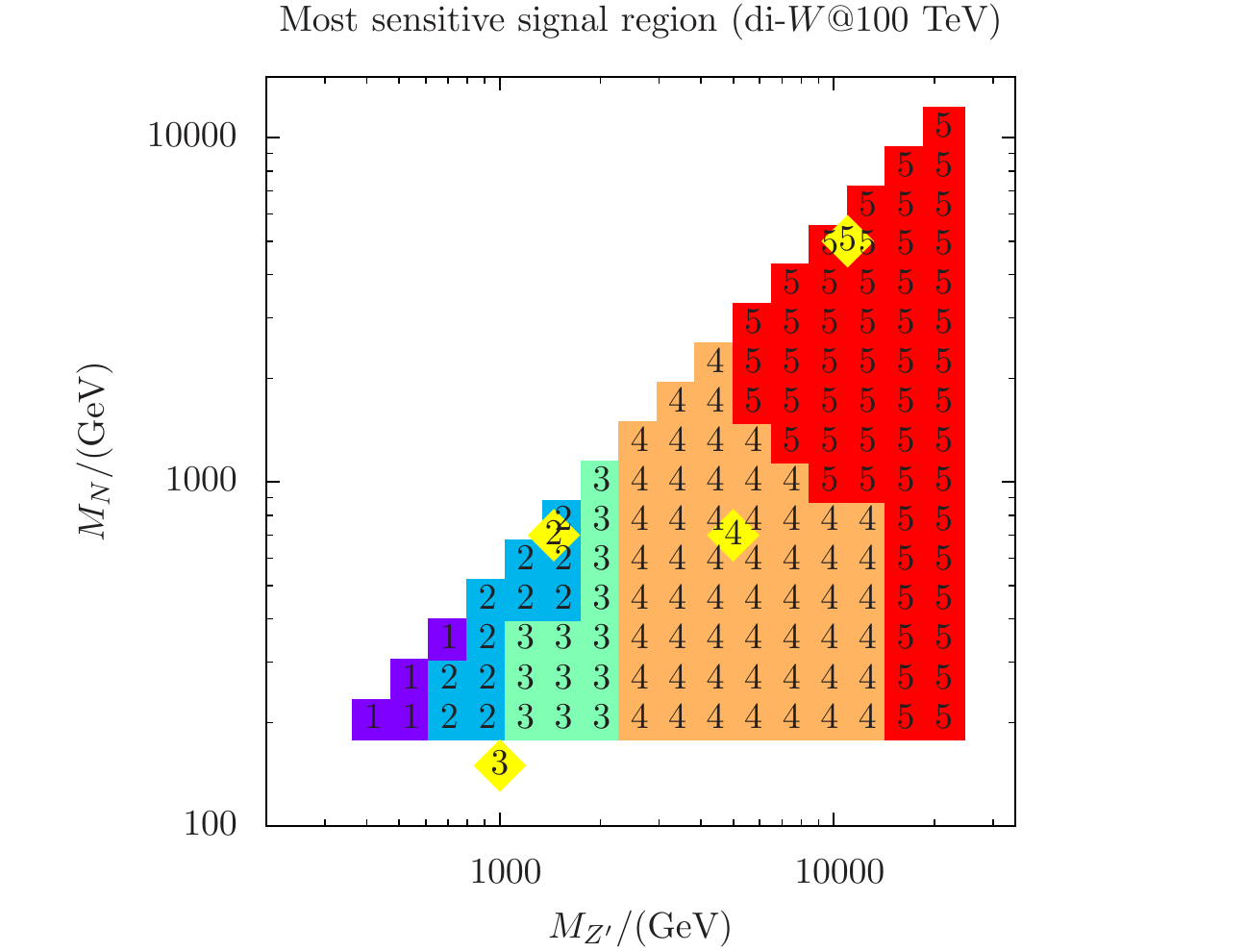}
\includegraphics[width=0.49\textwidth]{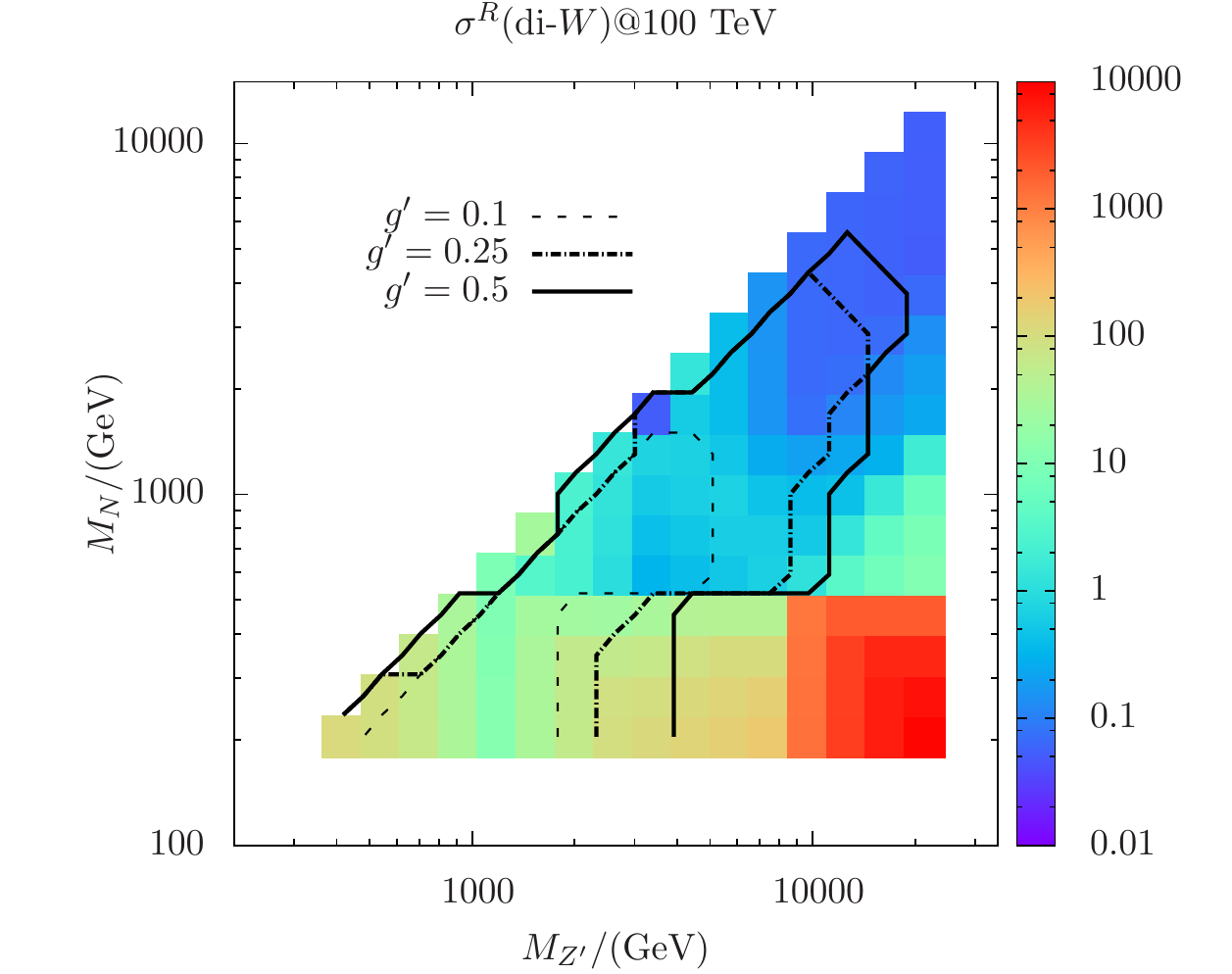} 
\end{center}
\caption{Left: the number on each grid shows the most sensitive signal region. 
Right: the color coding bar indicates the minimal cross section of di-$W$ channel (in fb) required for 3$\sigma$ signal significance. The curves give the discovery reaches of $Z'$ mediated RHN pair production with different $g'$. }
\label{wl14TeV} 
\end{figure}

 With the 4(5) signal regions that are optimized on 4(5) benchmark points for 14(100) TeV collision energy, we attempt to apply them to other girds on the $M_{Z'}-M_{{N}}$ plane to get a global outlook of the di-$W$ channel. The results are presented in Fig.~\ref{wl14TeV}.
The left panels give the most sensitive signal region on each grid; the numbers on the yellow diamonds correspond to the sequence number of the signal regions. We can see that the optimized searches on the benchmark points indeed provide the best sensitivity on their vicinities with similar kinematic properties. An exception occurs at the region with $M_{Z'} \gtrsim 5$ TeV and $M_{{N}} \lesssim 400$ GeV, where the PDF effect becomes important as we have discussed before.  Concretely, here the low energy RHN pair production via the off-shell $Z'$ dominates and thus the kinetic feature is more like S3 rather than S4. In the contrast, we do not see similar phenomena for the 100 TeV case since there the PDF effect is negligible.

On the right panels, we demonstrate the signal reaches (assuming 3000 fb$^{-1}$ luminosity) at 3$\sigma$ level using color codes. One may have some observations on the changes of search sensitivity when traveling on the $M_{Z'}-M_{{N}}$ plane. In the heavier RHN region (says with $M_{{N}} \sim 0.5 M_{Z'}$) where the PDF effect is not significant, the search sensitivities at both colliders improve as $M_{{N}}$ increases, since it results in harder final states. Similarly, in the lighter RHN region, says near 150 GeV, the sensitivities improve as increasing $M_{Z'}$ from 300 GeV to 1 TeV. But increasing the $M_{Z'}$ further, the sensitivities behave differently at two colliders. At the 100 TeV collider, increasing $M_{Z'}$ causes more and more serious lepton overlapping issue, which worsens the sensitivity substantially. To support that, in the lower right panel of Fig.~\ref{PDF} we show the distributions of the lepton number for different $M_{Z'}$, and we can see that the distributions do not change much  for $M_{Z'} \in [400,\,1000]$GeV, however, there is a significant drop as $M_{Z'}$ goes to 3 TeV; it become even worse for heavier $Z'$. The situation is different at 14 TeV, because for $M_{{N}}\sim 150$ GeV, as mentioned before, due to the PDF effect, a substantial fraction of RHNs are produced via low $x$ for $M_{Z'} \gtrsim 3$ TeV. So, the overlapping issue is relaxed and the search sensitivity is ameliorated for increasing $M_{Z'}$.

Now we interpret our results in the concrete model, the BLSM. We show the 3$\sigma$ exclusion limits for given parameter setups, $g'=0.1/0.25/0.5$. 
As we can see, the HL-LHC will be able to probe RHN mass up to about 2 TeV when $g' \gtrsim 0.2$; the light RHN and heavy $Z'$ corner, despite of a large signal rate, is still beyond exclusion owing to the lepton overlapping problem. As for the 100 TeV collider, it shows a remarkable enhancement in the RHN probing ability and even the heavy RHN region of $M_{{N}} \sim 5$ TeV can be covered.

 \subsection{The di-Higgs channel: boosted di-Higgs boson plus \ET}

In general, such a pure hadronic channel is not hopeful at the hadronic colliders, but the boosted Higgs bosons could provide a powerful tool for discrimination, particularly at the 100 TeV collider for the multi-TeV scale RHN. We will see that largely speaking this channel is the worst one among the three channels, except for a narrow region like $M_{{N}} \lesssim 500$ GeV and $M_{Z'} \gtrsim 4$ TeV, where the issue of lepton overlapping in the decay ${N}\ra W \ell$ is too severe. However, the di-Higgs channel still deserves a careful exploration from several aspects. First, the search strategy is absolutely different and the corresponding signature, boosted di-Higgs boson plus \ET, may be a generic sign of new physics~\cite{Kang:2015nga,Biswas:2016ffy,Basso:2015aee}. Second, it is a preparation for the $hW$ channel, which may turn out to be the best in some parameter space. Last but not least, this channel is lepton flavor independent and thus is complementary to other channels.

\subsubsection{Backgrounds and preselection with Higgs tagging}

The BGs mainly consist of the QCD multi-$b$-jets with $\ET$ due to the limited jet energy resolution, the semi-leptonic $t\bar{t}$ with leptons missed at the detector and $Zb\bar{b}$ with the subsequent decay $Z\to \nu \nu$; in the later two BGs, the other two bottom quarks come from the mis-identification of light quarks or gluons as $b$-jet. One can find the cross sections of BGs in Table~\ref{c1eff1}. More detailed discussions about the BGs can be found in Ref.~\cite{Kang:2015nga}, which shows that the irreducible QCD 4$b$-jets furnishes the dominant BG after applying all possible cuts. 
 \begin{table}[htb]
  \begin{tabular}{|c|c|c|c|c|c|c|c|c|c|c|c|c|}\hline
    & S2 & S3 & S4 &S5 & $t \bar{t}$ & $b\bar{b} b\bar{b}$ & $Z\bar{b} \bar{b}$  \\ \hline
  $\sigma_0$/pb &   1 & 1 & 1 & NO[1] & 803.4[29150]& 861[13530] & 109[1280]   \\
  $\sigma_{pre}/pb$ &  0.76[0.9] & 0.84[0.97] & 0.86[0.95] & NO[0.96] & 486.7[20120] & 398.9[6943] & 58.5[654.2]\\ \hline
  \end{tabular}  
   \caption{\label{c1eff1} Di-$h$ channel: Cross sections for signals and backgrounds before and after preselection. All the signal production cross sections have been normalized to 1 pb.}
\end{table}

Both at the 14 and 100 TeV colliders we impose four preselection cuts: 1) No lepton; 2) No $\tau$;  3) \ET $>$10 GeV; 4) At least two Higgs jet candidates. The loose $\ET$ cut is imposed to suppress the QCD background at the preselection level. In S1-S4 and S5, the Higgs bosons are normally and over boosted, respectively, which results in difference in the meanings thus tagging methods of a  ``Higgs jet". We address these differences in the following:

\begin{description}
\item[ Normally boosted] In this case the Higgs jet candidates is required to have substructure. The Delphes  EFlow  objects, in which the isolated leptons have been subtracted, are used for jet reconstruction. In the first, the fat-jets are reconstructed by the Cambridge/Aachen (C/A) algorithm~\cite{CA} with $R=1.4$. Then, the BDRS algorithm~\cite{Butterworth:2008iy} is applied on these fat-jets to resolve their substructures. Concretely speaking, a Higgs jet candidate should have a large mass drop and not too  asymmetric splitting during the declustering:
\begin{align}
\mu = \frac{m_{j_1}}{m_j} < 0.67, \quad y = \frac{\min (p^2_{T, j_1} , p^2_{T,j_2})}{m^2_j} \Delta R^2_{j_1, j_2} > 0.09.
\end{align}
Afterwards, the filtering method, i.e., the anti-kt jet algorithm with $R_{filt}=\min(0.3,R_{b\bar{b}}/2)$, is used to reconstruct the subjets inside each Higgs jet candidate and only the three hardest subjets are kept. Originally this step aims at suppressing the underlying events. Here, it has another function: a cone size $R=1.4$ may be already overlarge for some grids and thus the non-Higgs jets may contaminate the Higgs jet, and filtering helps to exclude them. At last, the rest of the EFlow  objects, i.e., those neither are isolated leptons nor belong to any Higgs jet candidates, are used for narrow (non-Higgs) jet reconstruction via the anti-kt algorithm with $R=0.4$; such jets are denoted as $j_{i}$. At the stage of preselection, we do not require $b$-tagging and mass conditions on Higgs jet.

\item[Over boosted] 
From the BDRS method used above, we can see that the jet substructure analysis is valid only if the angular separation between the two $b$-jets inside the Higgs jet is larger than $\sim$0.3, corresponding to $p_{T}(h) \lesssim 2 m_h /0.3  \sim 830$ GeV. Whereas $p_{T}(h)\gtrsim 2$ TeV for S5~\footnote{In fact, the resulting  angle separation is already near the LHC resolution limit $R_{min}\sim 0.2$~\cite{Spannowsky:2015eba,Bressler:2015uma}.} and thus there one cannot use the jet substructure analysis. Instead, the Higgs jet should be tagged as a whole by means of the anti-kt algorithm.  
In this algorithm, we scan over different values of $R \in [0.2,0.6]$ with step size 0.1. For each $R$ value, we count the number of the events of which the two leading jets invariant masses lie within [110,140]GeV and find that  $R=0.4$ can retain the largest number of events. Actually, $R$ takes 
0.4 both in the Higgs-tagging and normal jet reconstruction, so this value will be used for reconstructing all jets in S5; such jets are denoted as $j^{ak}$. After the preselection, we will impose the $b-$tagging condition on $j^{ak}$. 

\end{description}
For clearness, hereafter we will denote the Higgs jet candidates in the normally (over) boosted cases as $h^{ca}$ ($h^{ak}$). Note that at 100 TeV S5 is unlikely to have two $h^{ca}$, so in the preselection we only require at least two $h^{ak}$ for all benchmark points; for S2-S4, the cut of at least two $h^{ca}$ will be imposed after the preselection.

The cross sections of signal and BGs before and after preselection are given in Table~\ref{c1eff1}. For the 14 TeV case, the preselection reduces the number of BG events merely by a factor about $2$. As for signals, the selection efficiencies increase from S2 to S4,~\footnote{We do not make analysis on S1 because it gives non-boosted Higgs bosons with $p_T(h) < 200$ GeV (hence BDRS fails) and thus the search sensitivity is very low.} understood by nothing but the more and more boosted Higgs bosons thus the higher and higher BDRS Higgs-tagging efficiency. For the 100 TeV case, the signal event numbers, in particular for S3-S5, almost are not reduced after preselection, as is due to the loose preselection, namely requiring two $h^{ak}$ rather than two $h^{ca}$.

\subsubsection{Multivariable analysis}

In this channel, in addition to two Higgs jets plus \ET, we also include ISR jets in the final states; later we will explain the importance of this. Then, the complete variables list in the MVA are 
\begin{align}\label{14:hh}
&n_j, ~p_T(j_1), ~\ET, ~n_h, ~m_{T_2}(h^{ca}_1,h^{ca}_2), \nonumber \\
&p_T(h^{ca}_{1,2}), ~\phi(h^{ca}_{1,2}), ~m(h^{ca}_{1,2}),~\eta(h^{ca}_{1,2}); \\ \label{100:hh}
&n_j, ~p_T(j_1), ~\ET, ~n_h, ~m_{T_2}(h^{ca}_1,h^{ca}_2), ~p_T(h^{ca}_{1,2}), ~m(h^{ca}_{1,2}),
 \nonumber
  \\
&p_T(j^{ak}_1),  ~\Delta \phi(h^{ak}_1, \vec{p}^{\text{miss}}), ~p_T(h^{ak}_{1,2}), ~m(h^{ak}_{1,2}),
\end{align}
where $p_T(j_1)$ is the transverse momentum of the leading non-Higgs jet. The elements in Eq.~(\ref{14:hh}) and Eq.~(\ref{100:hh}) are for the 14 and 100 TeV cases, respectively. At the 14 TeV LHC, the full 4-momentum information of the Higgs jets would be very helpful and the corresponding variables are incorporated in the second line of Eq.~(\ref{14:hh}). Although relatively less powerful, we still include $m_{T_2}(h^{ca}_1,h^{ca}_2)=   \min_{\slashed{p}_T^1 + \slashed{p}_T^2 = \ET } \ [  \max(m_T(h^{ca}, \slashed{p}_T^1),m_T(h^{ca}, \slashed{p}_T^2) ) \ ]$ in the MVA. At the 100 TeV collider, to improve the analysis for S5 we add quite a few additional variables in the second line of Eq.~(\ref{100:hh}), with all jets labelled with superscript ``ak" to indicate the different preselection than the counterparts for S1-S4; see the discussions before.

 \begin{table}[htb]
  \begin{tabular}{|c|c|c|c|c|}\hline
   S2 & S3 & S4 &S5\\ \hline
  $\ET [\Delta \phi(j^{ak}_1,\vec p^{miss})]$  & $\eta(j^{ca}_{1,2})[\Delta \phi(j^{ak}_1,\vec p^{miss})]$ & $\eta(h^{ca}_{1,2})[\Delta \phi(j^{ak}_1,\vec p^{miss})]$ & NO$[\Delta \phi(h^{ak}_1,\vec p^{miss})]$ \\
  $\phi(h^{ca}_{1,2})[\ET]$ & $\phi(h^{ca}_{1,2})[m({h^{ca}_1})]$ & $\phi(h^{ca}_{1,2})[\ET]$ &NO$[m(h^{ak}_1)]$\\
  $m({h^{ca}_1})[m({h^{ca}_1})]$ & $p_T(h^{ca}_1)[p_T(h^{ak}_1)]$ & $\ET [m({h^{ca}_1})]$ & NO$[n_j]$ \\
  $\eta(h^{ca}_{1,2})[m({h^{ca}_2})]$ & $p_T(j_1)[n_j]$ & $m({h^{ca}_1})[n_j]$ & NO$[p_T(j_1)]$ \\
  $m({h^{ca}_2})[n_j]$ & $m({h^{ca}_2})[\ET]$ & $n_j[m({h^{ca}_2})]$ & NO$[m({h^{ak}_2})]$\\ \hline
  \end{tabular}
   \caption{Di-$h$ channel: top-5 variables in BDT analysis. 
   \label{t5:hh}}
\end{table}
Table~\ref{t5:hh} gives the five most important variables in the BDT analysis. $\Delta \phi(j^{ak}_1,\vec p^{miss})$, the azimuthal angle difference between the leading jet and \ET, plays a remarkable role for the 100 TeV case. This has a clear explanation. The \ET of BGs comes from jet energy mis-measurement and thus is supposed to closely follow the  direction of the leading jet. 
As a result $\Delta \phi(j^{ak}_1,\vec p^{miss})$ of the signal process is much larger than that of BGs. Actually, for the 14 TeV case, the similar information has been encoded in $\phi/\eta(h_{1,2}^{ca})$, making it remarkable there. The \ET, masses of Higgs jets and  the number of non-Higgs jets are also very important. Compared to BGs, for signals the \ET and masses of the Higgs jets candidate are much larger, while the non-Higgs jets are much fewer. The final results after training the BDT are shown in Table~\ref{c1res}.

 \begin{table}[htb]
  \begin{tabular}{|c|c|c|c|c|c|c|c|c|c|c|c|}\hline
   &  Cut & $\epsilon$(SIG) & $\sigma$(BG)(fb) & signal reach(fb)@ 3000 fb$^{-1}$ \\ \hline
S2 & BDT$>$ 0.2[0.2] & 0.015[0.0078] & 0.012[0.067] & 0.42[2.2] \\ \hline
S3 & BDT$>$0.2[0.2] & 0.01[0.0045] & 0.39[2.01] & 6.8[69.2] \\ \hline
S4 & BDT$>$0.2[0.2] & 0.048[0.029] & 0.034[0.067] &  0.24[0.6] \\ \hline
S5 & BDT$>$ NO[0.3] &NO[0.028] & NO[2.73]  &  NO[15.0]\\ \hline
  \end{tabular}
      \caption{\label{c1res} Di-$h$ channel: cuts efficiencies and signal reaches.}
\end{table}

S2 and S4, which give normally boosted Higgs boson pair, are the most sensitive signal regions, with S4 mildly better due to the harder final states from the heavier $M_{Z'}$ decay. The signal reach of S3 is much weaker than those of S2 and S4, even though all of them give similarly energetic Higgs bosons.  
The reason is mainly due to the smaller \ET of S3.  In the decay ${N} \to h \nu_L$ most of the energy of RHN is carried away by the Higgs boson, because RHN and $h$ have similar masses while $\nu_L$ is much lighter. In addition to that, both $\nu_L$ and $h$ are boosted along the RHN flying direction in the lab frame, which gives rise to a cancelation between the two $\nu_L$ momentums; moreover, it renders a small $\Delta \phi(j^{ak}_1,\vec p^{miss})$, thus further weakening the signal reach of S3. At 100 TeV, the search sensitivity of S5 is also much ($\sim$ one order of magnitude) weaker than those of S2 and S4, because the number of background events  increases dramatically in the absence of the requirement of jet-substructure.

 \subsubsection{Results and analysis}

\begin{figure}[htb]
\begin{center}
\includegraphics[width=0.48\textwidth]{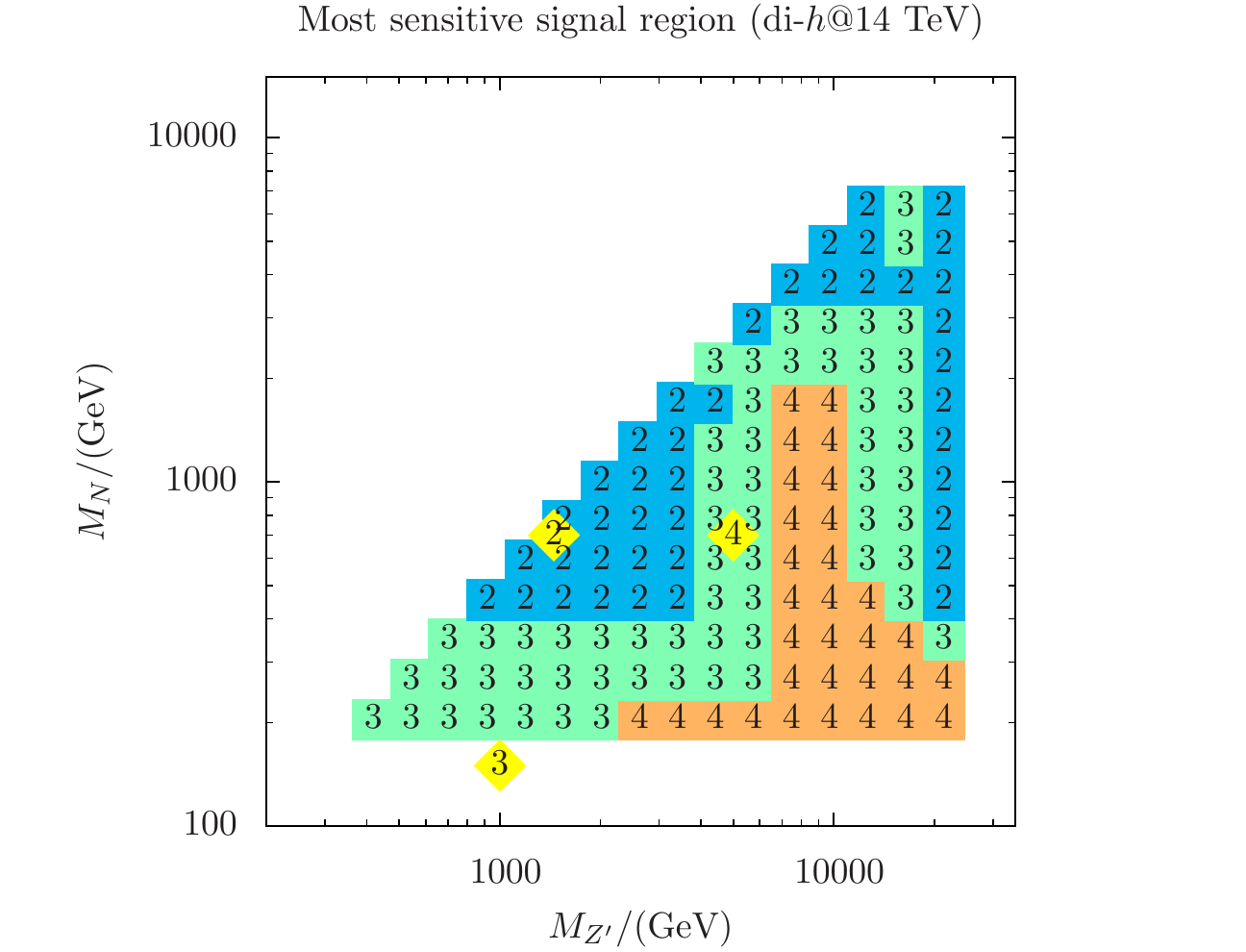}
\includegraphics[width=0.48\textwidth]{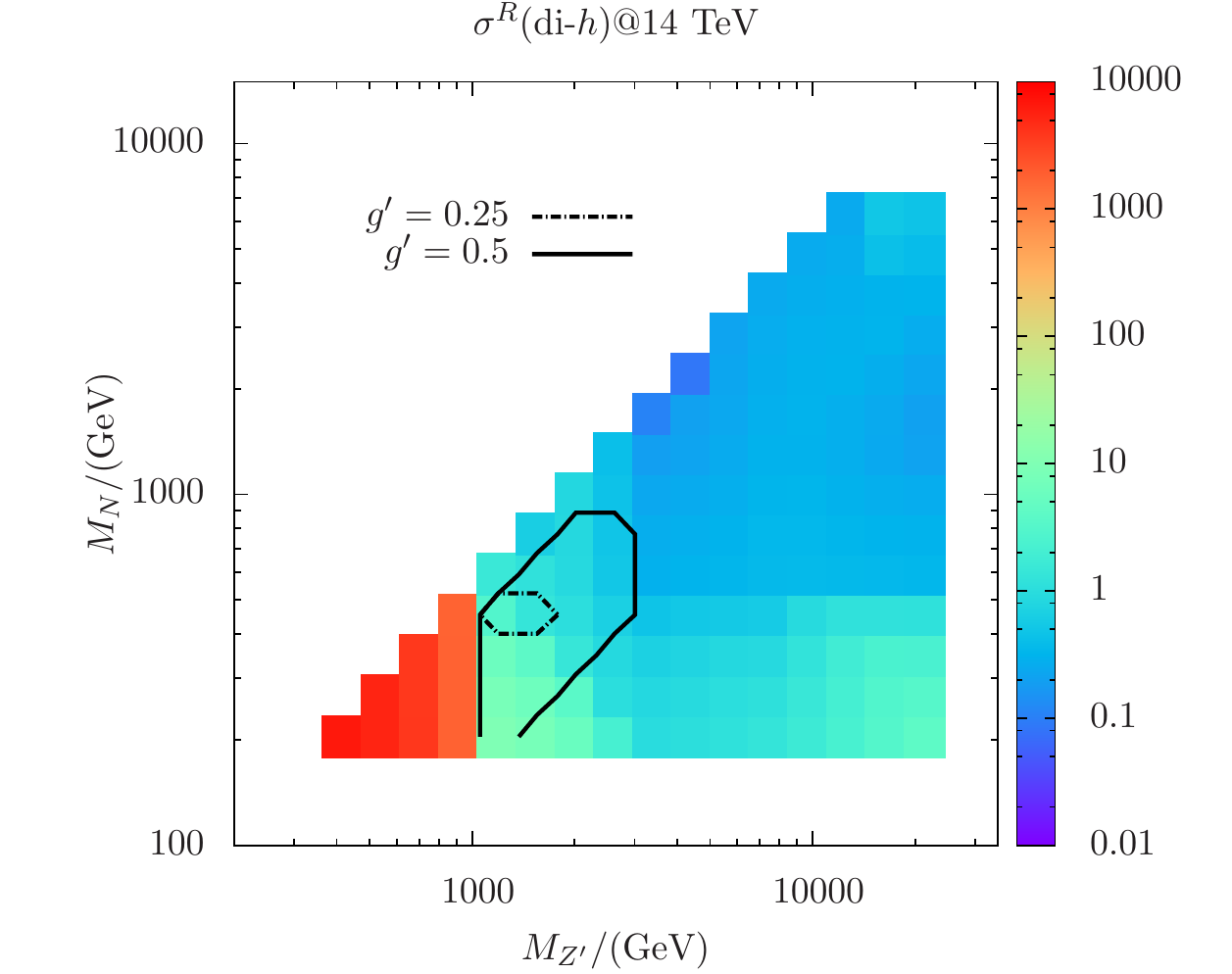} 
\includegraphics[width=0.49\textwidth]{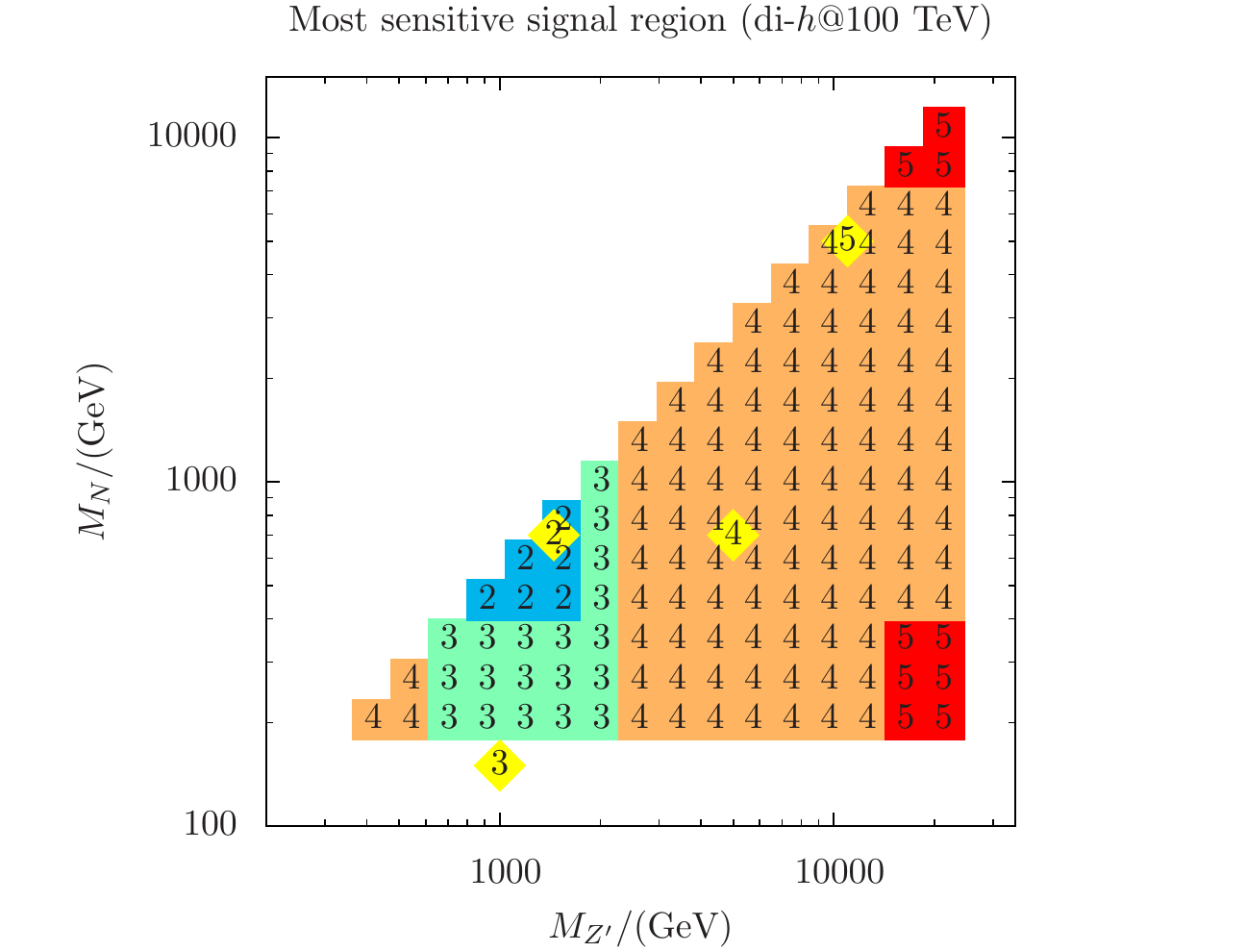}
\includegraphics[width=0.48\textwidth]{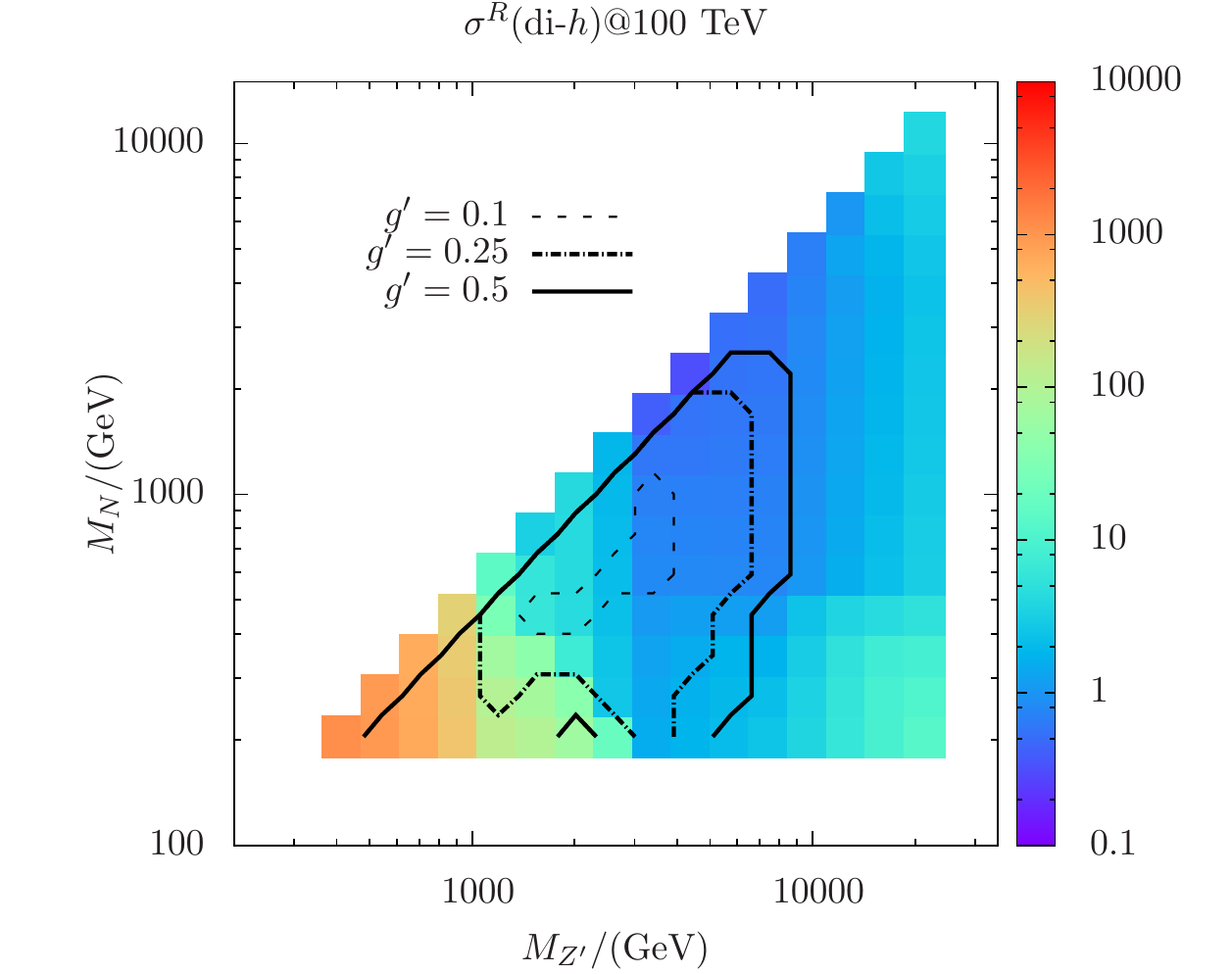} 
\end{center}
\caption{Same as Fig~\ref{wl14TeV} but for di-$h$ channel. }
\label{hh:result} 
\end{figure}
\begin{figure}[htb]
  \centering
  \includegraphics[width=0.5\textwidth]{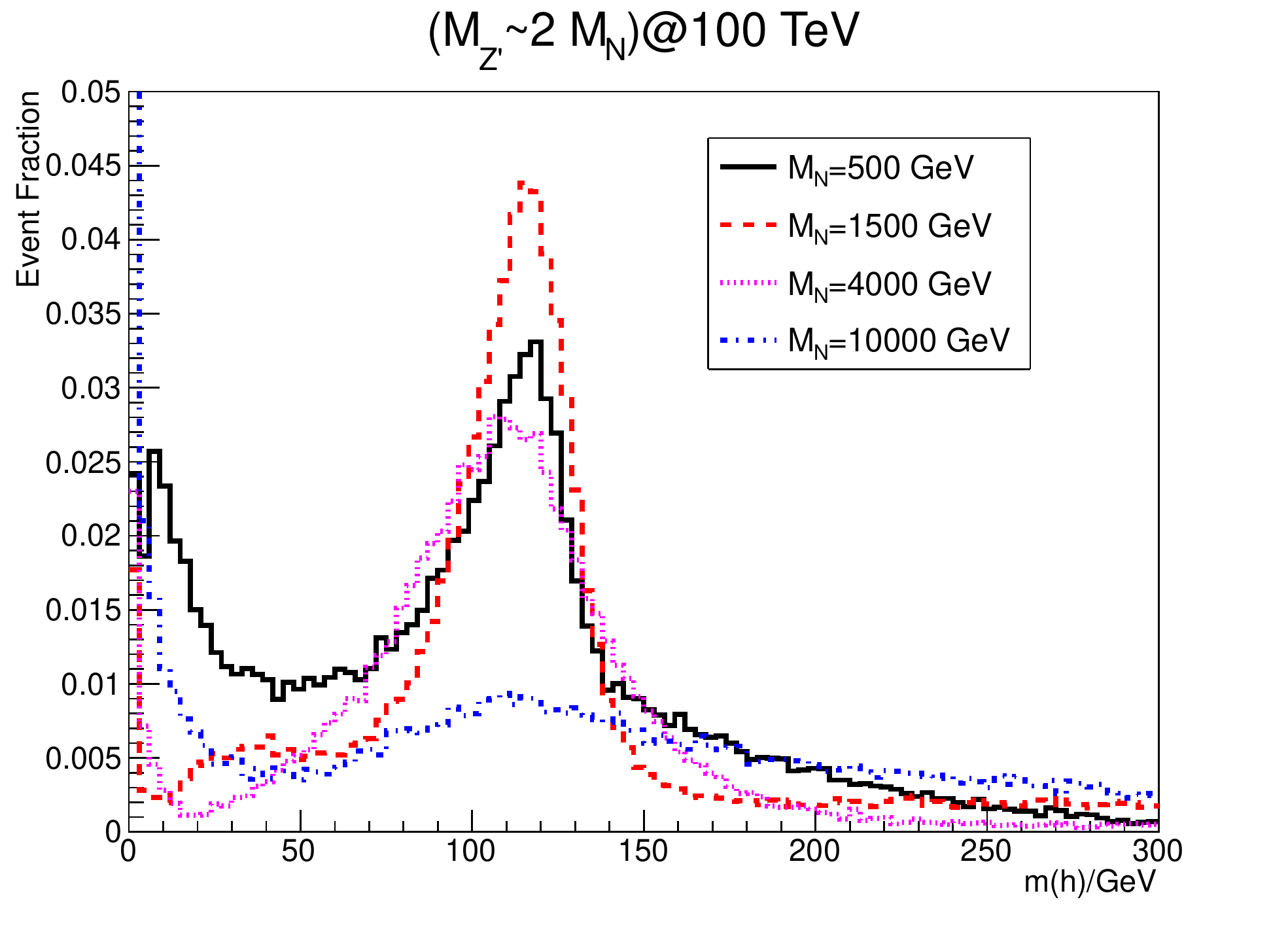}
  \caption{\label{exbost} The distributions of the leading Higgs tagged jet invariant mass for different $M_N$.}
  \label{BDRS:h}
\end{figure}

Again, we apply our analyses that are optimized on the benchmark points to other grids on the $M_{Z'}$-$M_{{N}}$ plane. The search results are displayed in Fig.~\ref{hh:result}. 
From it we make a long list of observations. 
\begin{itemize}
\item At 14 TeV, S2 and S3 indeed give the best signal reaches in their vicinities. But in the heavy ${Z'}$ and RHN region, namely around S4 all signal regions almost give similar sensitivities (might as a result of PDF effect); see the top right panel. Consequently, the fluctuation leads to a random distribution of the most sensitive signal region. 

\item  In practice S4 gives the most sensitive search for the benchmark point 5. 
This tells us that for this benchmark point, tagging the two $h^{ca}$ in the final state is crucial to improve the signal significance.
S5 is more suitable for the even heavier mass region, e.g., $M_{{N}} \gtrsim 7$ TeV, where the number of signal events that contain two $h^{ca}$ is too small. 

\item In the right panels of Fig.~\ref{hh:result}, one can see that decreasing $M_{{N}}$ with fixed $M_{Z'}$ leads to worse  search sensitivity. This is because the smaller mass splitting between RHN and Higgs renders the softer and more collinear $\nu_L$-pair, i.e., the smaller $\ET$ and $\Delta \phi(j^{ak}_1,\vec p^{miss})$, as we have explained before. 

\item At 100 TeV, fixing $M_{{N}}$ while increasing $M_{Z'}$, the search sensitive goes through an improvement then deterioration. This is mainly due to the over boosted effect on the Higgs boson. To show this, in Fig.~\ref{BDRS:h} we show the invariant mass distribution of the leading BDRS Higgs jet for different RHN mass at 100 TeV. We can see that, after imposing the Higgs mass condition on the BDRS jet, says $\in [100,150]$ GeV, the model with $M_{{N}}=1500$ GeV has the highest Higgs-tagging efficiency; an even heavier RHN would begin to over boost the Higgs boson. This explains why the signal reach is best for $M_{Z'} \sim 3$ TeV. On the contrary, at 14 TeV the reduction of the search sensitivity while increasing $M_{Z'}$ is much milder than that at 100 TeV, ascribed to the PDF effect.

\item In the heavy $Z'$ and light RHN region, the di-$h$ and di-$W$ channels are suffering the problems of small $\ET$ and overlapping leptons, respectively. Comparing the right panels of Fig.~\ref{} and Fig.~\ref{hh:result}, it is seen that the latter problem starts from a smaller $M_{Z'}$. This means that  in the region $M_{Z'}\gtrsim2 $TeV and $M_{{N}}\lesssim0.5 $TeV, the di-$h$ channel could provide a better sensitivity than that of the di-$W$ channel.

\end{itemize}

Note that our analysis of the di-Higgs channel here is not as promising as the one employed in Ref.~\cite{Kang:2015nga} especially in the region of $M_{{N}}\lesssim500$ GeV. Because there the substructure analysis is more refined, i.e. the cone size parameter in jet reconstruction is optimized at different $M_{{N}}$ to gain the maximal sensitivity. Here we are exploring a much wider two-dimensional plane with varying $M_{Z'}$ but at the price of optimization; new interesting phenomena at collider such as the over boosted Higgs boson arise here.

 \subsection{The $hW$-channel at the 14 \& 100 TeV colliders}

In this subsection, we consider the channel with one RHN decaying as $\ell W(\ra \nu_L \ell)$ and the other decaying as $\nu_L h(\ra b\bar b)$. The leptonic $W$ helps to suppress the huge $t\bar{t}$ and $Wb\bar{b}$ BGs; the $b\bar b$ mode of Higgs decay enables us to adopt the BDRS jet-substructure analysis. This mixed channel  take advantages of a smaller background for $W\ell$ channel and less sensitive to RHN boost for $hh$ channel. So it will provide the best signal reach in some parameter space. 

\subsubsection{BGs and preselection}

The fully leptonic $t\bar{t}$ ($t_\ell \bar{t}_\ell$)  and $Z/\gamma[\to \ell\bar{\ell}] b\bar{b}$ constitute the main BGs for this channel. In the monte carlo generation of the $t\bar{t}$ events, we let top quarks decay at the matrix element level. For the $Z/\gamma[\to \ell \bar{\ell}] b\bar{b}$ background event generation, we impose the cut $p_T(\ell) > 20$ GeV to suppress the huge contribution from the $\gamma$ mediated processes. At 14 TeV, the production cross section of $Z/\gamma[\to \ell \bar{\ell}] b\bar{b}$ is around one order of magnitude smaller than that of $t_\ell \bar{t}_\ell$; moreover, increasing the collision energy from 14 TeV to 100 TeV the latter BG shows a much larger magnitude of the increase of cross section. As for the $t\bar{t}V$ BGs, their contributions are $\sim{\cal O}(1\%)$. Nevertheless, they might still be important for the later analysis, because they have large \ET and complicated final states, which hamper the distinguishability from signals.

At both the 14 and 100 TeV colliders the preselection requires A) at least two isolated leptons and B) at least one Higgs jet candidate. The requirement of Higgs jet candidate is the same as in the di-Higgs channel, i.e., $j^{ak}$ for S5 at 100 TeV. The preselection efficiency for signal and BGs are shown in Table~\ref{c2eff1}. Like in the di-$h$ channel, we will not take S1 into consideration here. As we can learn from results of the di-$W$ channel and di-$h$ channel, the preselection efficiencies of the signal processes are mainly controlled by the number of lepton cut, while the Higgs jet number cut is much looser. So the efficiencies ranked similarly as in the di-$W$ channel, i.e., $S3<S2<S4(<S5)$.

\begin{table}[htb]
  \begin{tabular}{|c|c|c|c|c|c|c|c|c|c|c|c|c}\hline
    & S2 & S3 & S4&S5 & $t_l \bar{t}_l$ & $t\bar{t} Z $ & $t \bar{t} W$ & $Z/\gamma[\to \ell\bar\ell] b \bar{b}$ \\ \hline
  $\sigma_0$/pb &  1 & 1 & 1&1 & 36.4[1322.5] & 0.93[55] & 0.67[16.6]& 4.4[71.1]  \\
  $\sigma_{pre}/pb$ & 0.12[0.085] & 0.10[0.067] & 0.15[0.14]&NO[0.16] & 10.8[284.6] & 0.038[1.7] & 0.026[0.39] & 1.46[28.8]\\ \hline
  \end{tabular}
   \caption{\label{c2eff1} $hW$ channel: Cross sections for signals and backgrounds before and after preselection. All the signal production cross sections have been normalized to 1 pb.}
   \label{hwpre}
\end{table}

\subsubsection{Multivariable analysis and results}

In the $hW$ channel, we use the following variables in MVA:
\begin{align}
&n_j, ~p_T(\ell_1), ~p_T(j_1), ~E^{\text{miss}}_T, ~\Delta r (\ell_1,\ell_2), ~m(\ell_1,\ell_2), \nonumber \\ 
&~\Delta \phi (\ell_1, \vec{p}^{\text{miss}}), ~\Delta \phi (j_1, \vec{p}^{\text{miss}}), ~\Delta \phi(\ell_1,j_1), ~m(j_1,j_2),  \nonumber \\
&m(h), ~p_T(h), ~\Delta \phi (h,\vec{p}^{\text{miss}} ), ~\Delta r(\ell_1, h), ~m_{T_2} (h,\ell \ell) ; \\
& p_T(h^{ak}_2), ~m(h^{ak}_1), ~m_{T_2} (\ell \ell,h^{ak}_1), ~\Delta \phi (h^{ak}_1, \vec{p}^{\text{miss}} ) .
\end{align}
Again, elements in the last line are only for the 100 TeV case. We use a large number of angular variables, because there are strong angular correlations between the final states of the signal processes, especially in the heavy $Z'$ and light RHN region. In the signal process, the dilepton invariant mass $m(\ell_1,\ell_2)$ shows a kinematic edge at $ \L{M^2_{{N}}- m^2_{W}}\R^{1/2}$, which can be used to suppress the BGs where the lepton pair is from $Z$ decay. Here the $m_{T_2}(h,\ell \ell)$ variable is constructed using the Higgs boson and the di-lepton from the RHNs decay. In order to tag the over boosted Higgs boson in S5, we adopt a similar method used in the di-$h$ channel; the Higgs jets are denoted as $h^{ak}$ as before.

The top five variables in the BDT analyses are listed in Table~\ref{5:wh}. For S2, RHN is relatively heavy thus giving a large \ET, which plays a very important role. For S2 at 100 TeV, there is a substantial fraction of the events having large energy, and hence the two leptons, Higgs boson and \ET tend to collimate with each other. Then, the angular variables become important. For S3, the RHN is fairly light, so $\nu_L$ can only carry away a small fraction of the energy of ${N}$; see the arguments in the di-$h$ channel. S4 shares partial features both with S2 and S3, and both $\ET$ and angular variables come to contribute. As for S5, $m_{\ell \ell}$ is the most remarkable variable, followed by the invariant mass of $h^{ak}$.

\begin{table}[htp]
\begin{tabular}{|c|c|c|c|}\hline
  S2 & S3 & S4 & S5\\ \hline
  $E^{miss}_T[\Delta\phi(\ell,j)]$ & $[\Delta r(\ell,\ell)][\Delta r(\ell,\ell)]$ & $[\Delta r(\ell,\ell)][\Delta r(\ell,\ell)]$ & NO$[m_{\ell\ell}]$ \\
  $p_T(\ell_1)[[\Delta r(\ell,h^{ca})]]$ & $\Delta \phi (\ell,\ell)[\Delta \phi(\ell,\vec p^{miss})]$ & $m_{T_2}(h,\ell \ell)[m_{T_2}(\ell\ell,h^{ca}j)]$ & NO$[m(h^{ak})$\\
  $\Delta \phi(\ell,j)[\ET]$ & $\Delta \phi(\ell,\vec{p}^{miss})[[\Delta r(\ell,h^{ak}_1)]]$ & $E^{miss}_T[\ET]$ &NO$[\Delta \phi(\ell,\vec p^{miss})]$ \\
  $\Delta \phi(\ell,\vec{p}^{miss})[\Delta r(\ell,\ell)]$ & $\Delta r (\ell,h)[\Delta \phi(h^{ca},\vec p^{miss})]$ & $\Delta r (\ell,h)[\Delta \phi(\ell,\vec p^{miss})]$ & NO$[\Delta r(\ell,h^{ak}_1)]$ \\
  $\Delta r (\ell,h)[\Delta \phi(\ell,\vec p^{miss})]$ & $m_{\ell\ell}[\Delta \phi(j_1,\vec p^{miss})]$ & $\Delta \phi(\ell,\ell)[m_{\ell\ell}]$ & NO$[\Delta \phi(\ell,j)]$\\ \hline
\end{tabular}
\caption{$hW$-channel: top 5 variables in BDT anslysis.}
\label{5:wh}
\end{table}%

 After imposing cuts on the BDT response for signal and BGs in each signal region, the cut efficiency for the signal and  cross sections of remaining BGs are given in Table~\ref{cuts:wh}.  As the other two channels, the search sensitivities in the signal regions rank as $S3<S2<S4$. Like the di-$h$ channel, the search sensitivity in $S5$, in spite of its hard final states, is even worse than that in $S2$. This fact again supports that the BGs would grow rapidly without the BDRS Higgs tagging. Typically, for each signal region the corresponding search sensitivity of the $hW$ channel lies between those of the di-$h$ and di-$W$ channels. However, as we will see later, it may be the best one in some specific mass regions. 
 \begin{table}[htp]
 \begin{tabular}{|c|c|c|c|c|} \hline
   &  Cut & $\epsilon$[SIG] & $\sigma$[BG][fb] & signal reach[fb]@ 3000 fb$^{-1}$ \\ \hline
SR2 & BDT$>$ 0.2[0.2] & 0.02[0.0075] & 0.001 [0.015]& 0.087[0.94] \\ \hline
SR3 & BDT$>$0.2[0.15] & 0.013[0.0063]& 0.0076[0.49] & 0.38[13.2]\\ \hline
SR4 & BDT$>$0.2[0.2] & 0.038[0.029] & $2.5 \times 10^{-4}[0.082]$ & 0.023[0.18] \\ \hline
 SR5 & BDT$>$NO[0.3] & NO[0.055] & NO[0.47] & NO[1.4] \\ \hline
 \end{tabular}
\caption{$hW$ channel: cuts efficiencies and signal reaches.}

\label{cuts:wh}
\end{table}%

\begin{figure}[htb]
\begin{center}
\includegraphics[width=0.48\textwidth]{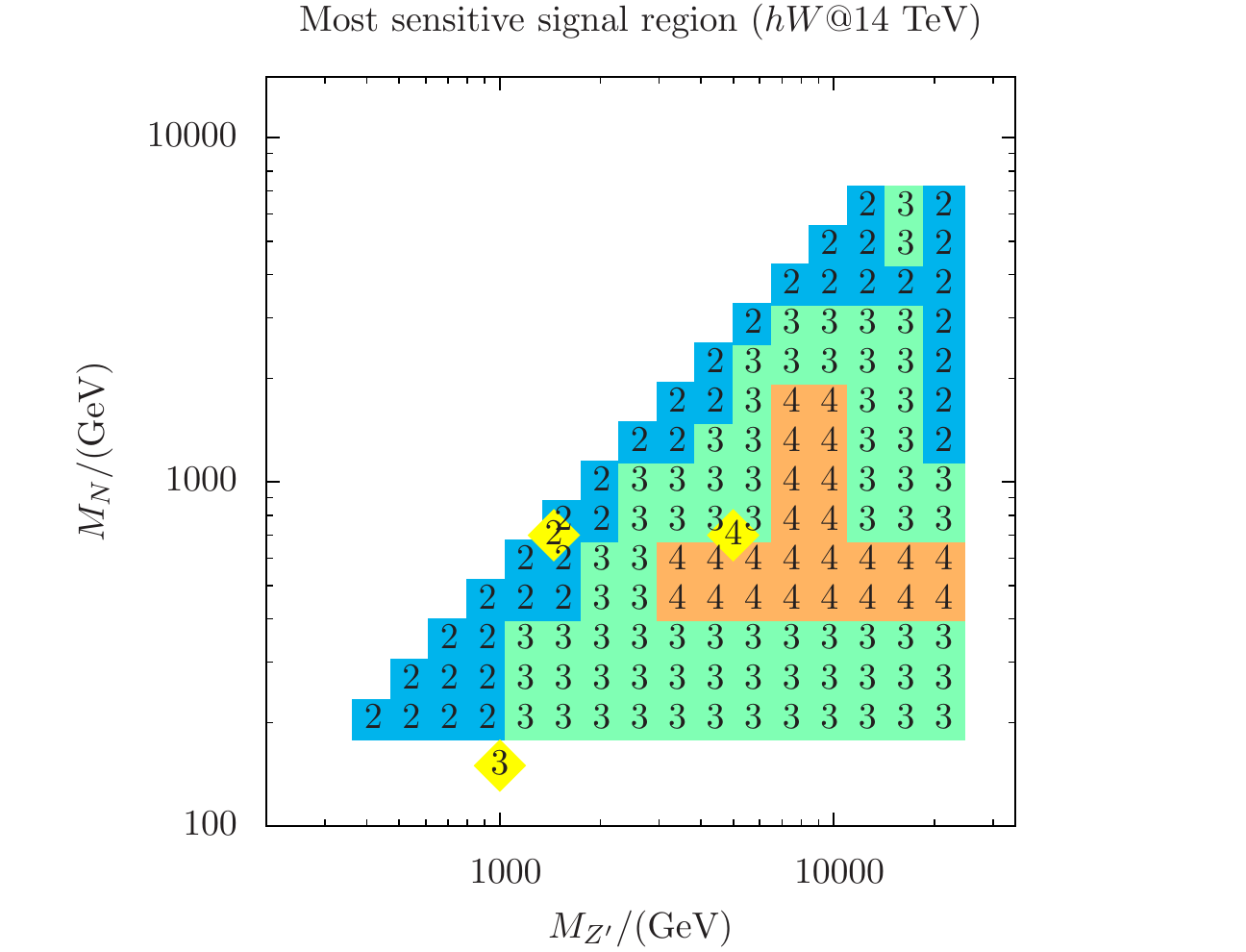}
\includegraphics[width=0.48\textwidth]{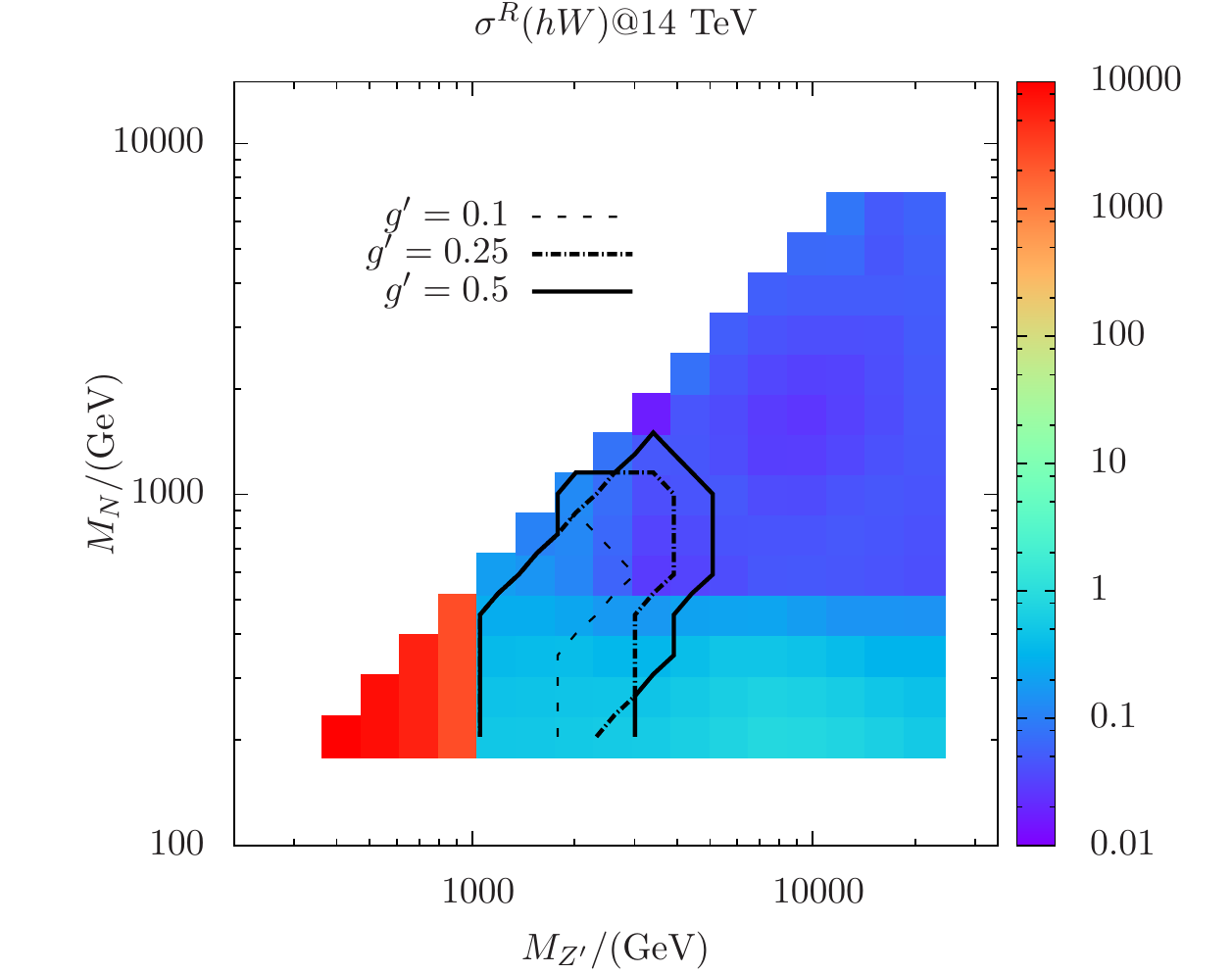} 
\includegraphics[width=0.48\textwidth]{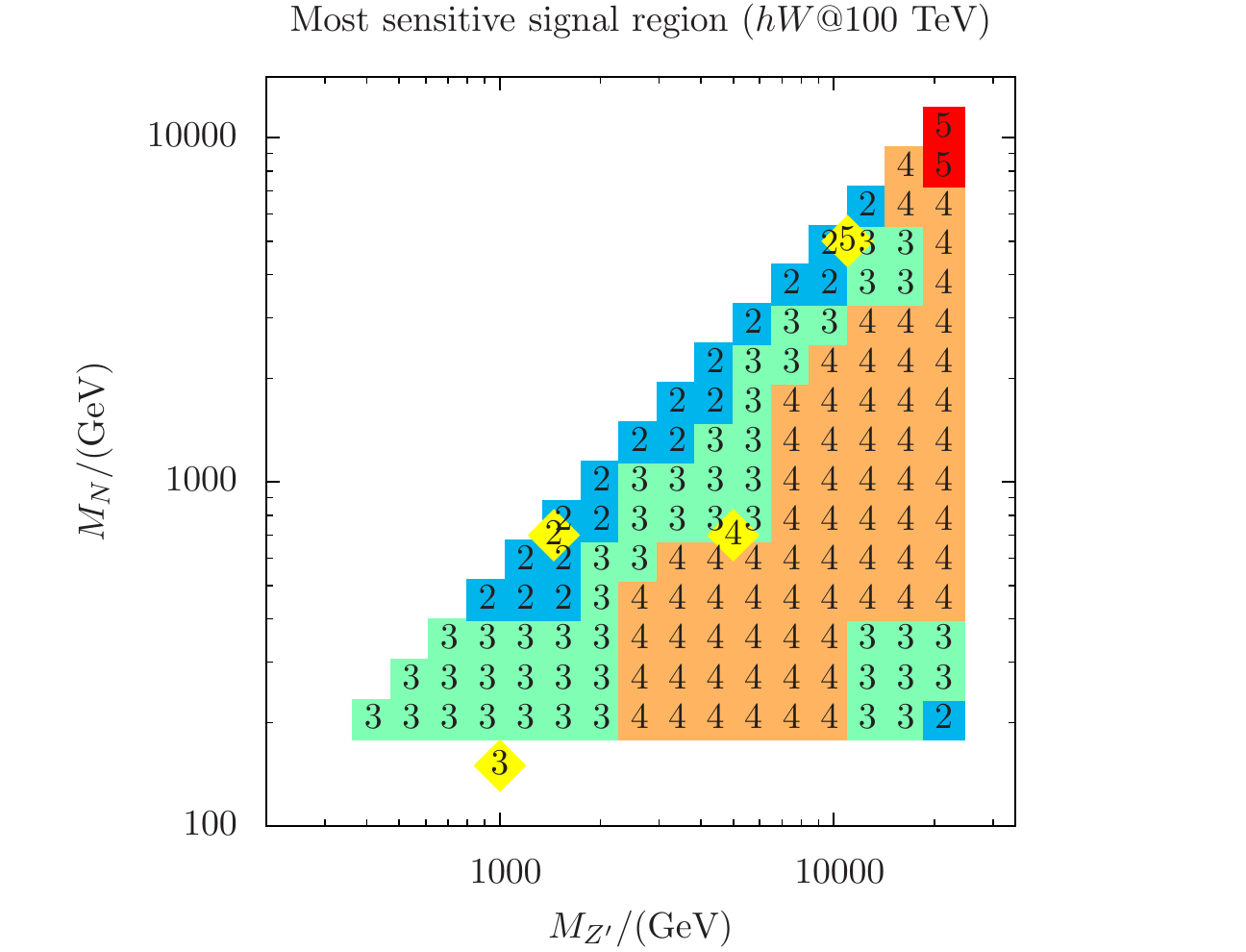}
\includegraphics[width=0.48\textwidth]{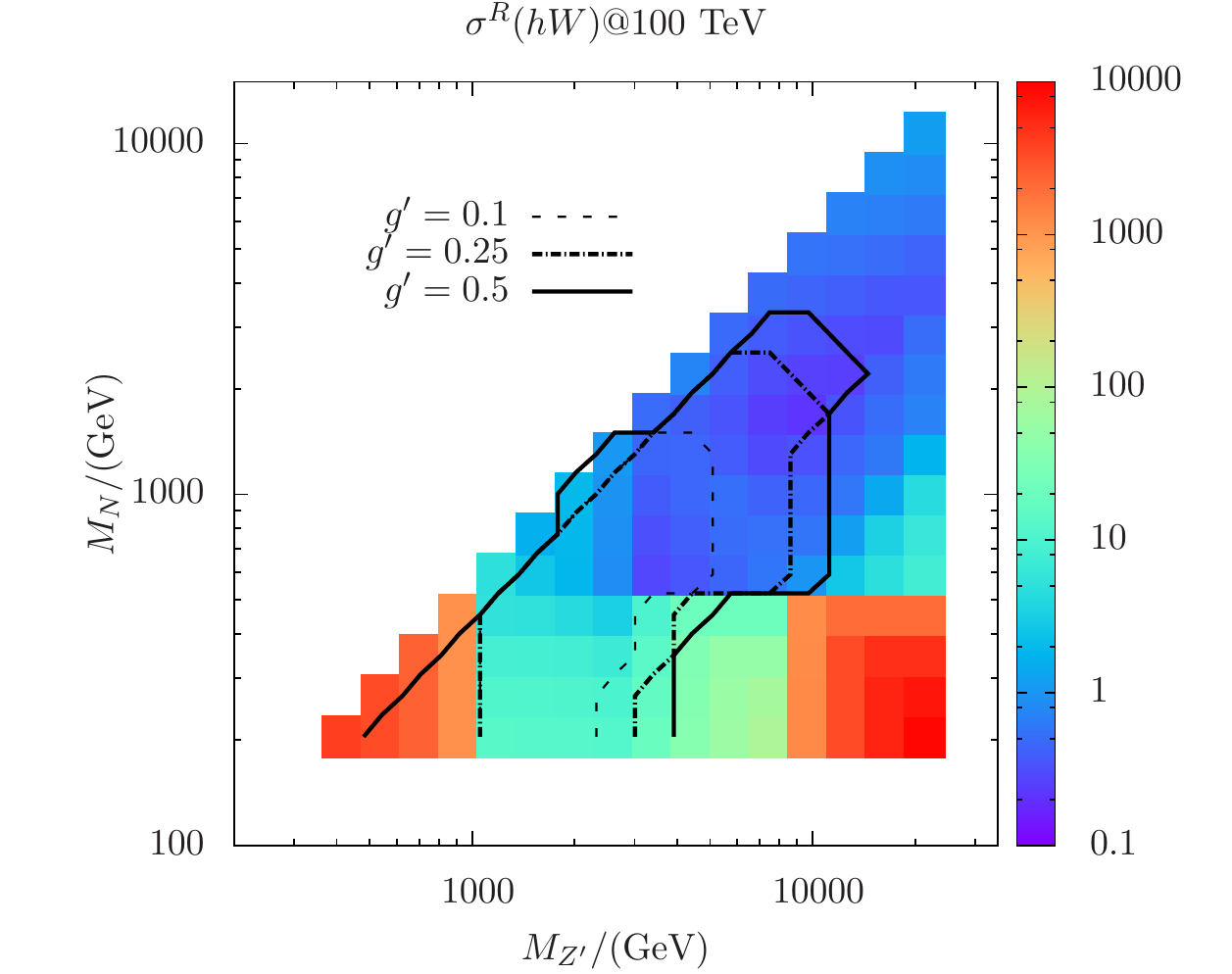} 
\end{center}
\caption{Same as Fig~\ref{wl14TeV} but for $hW$ channel.}
\label{wh:result} 
\end{figure}

On the gridded $M_{Z'}-M_{{N}}$ plane, the distributions of the best signal region and the signal reach are displayed in Fig.~\ref{wh:result}. We can understand the overall results of this channel based on the inherited features from the di-$h$ and di-$W$ channels, so we do not repeat the descriptions of the features exhibited in the figures and the corresponding explanations. Here we just want to stress that the $hW$ channel could provide the strongest search sensitive in the light RHN but relatively heavy $Z'$ region, where it on the one hand  is not subjected to lepton overlapping as in the di-$W$ channel and on the other hand has a small BG in comparison to the di-$h$ channel.


\subsection{Searching for RHN pair in BLSM with combined channels}

\begin{figure}[htb]
\begin{center}
\includegraphics[width=0.48\textwidth]{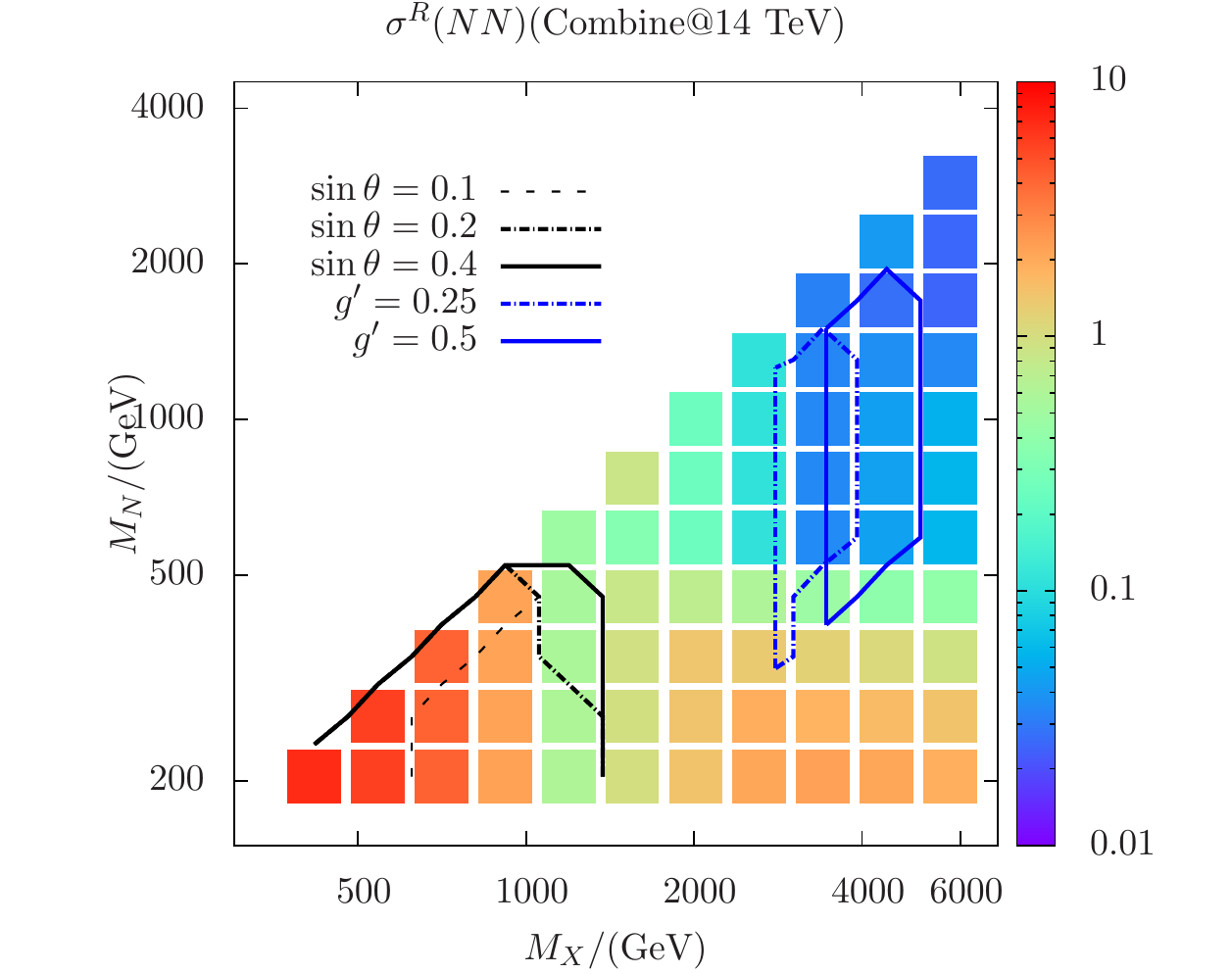}
\includegraphics[width=0.49\textwidth]{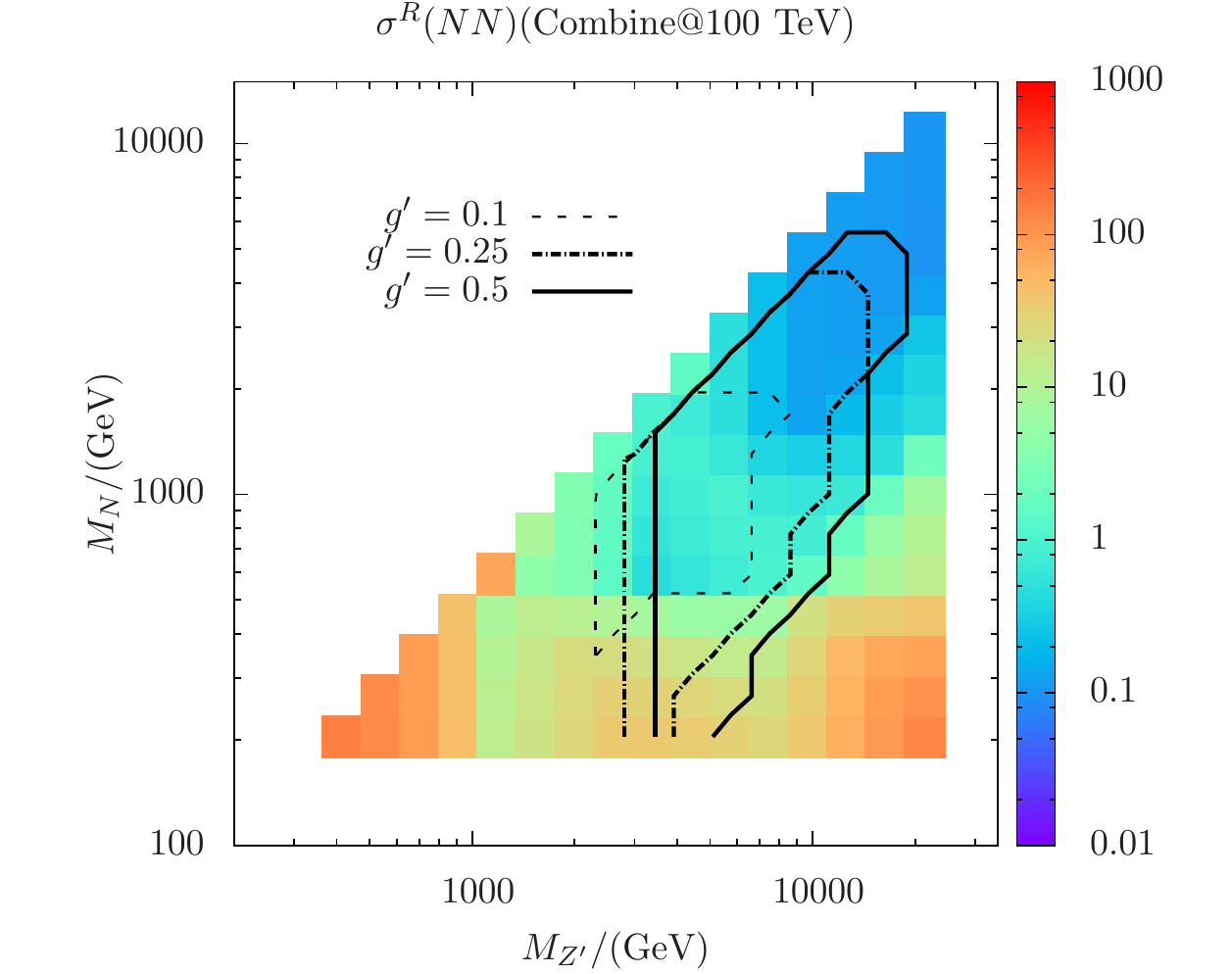} 
\end{center}
\caption{The color coding bar indicates the minimal cross section of ${N}$ pair production(in fb) required for 3$\sigma$ signal significance after combining searches for all three signatures. The discovery reaches for Higgs/Z' mediated $N$ pair production with different parameter setups are also shown.}
\label{comb:result} 
\end{figure}

Having employed analysis for the individual channel, in this subsection we would like to apply all of them to the benchmark model, BLSM (whose details can be found in Appendix.~\ref{spectra}), to investigate the best prospect of RHN pair search in this model. 
The combined $3\sigma $ level signal reach can be estimated by
\begin{align}
\sqrt{\L \frac{\sigma_{tot} ~ \text{Br}(\text{di-}h) \cdot \epsilon_s(\text{di-}h) \mathcal{L} }{\sqrt{B(\text{di-}h) + (0.05 B(\text{di-}h))^2}}  \R^2+(\text{di-}h \to \text{di-}W)+(\text{di-}h \to h W)} =3, \label{eq:comb1}
\end{align}
where $\sigma_{tot}$ is the RHN pair production cross section; Br$(\text{di-}h)$ is the branching ratio of the RHN pair decay into di-$h$; $\epsilon_s(\text{di-}h)$ and $B(\text{di-}h)$ respectively are the signal efficiency and number of background events obtained before. By using the $3\sigma$ signal reach of each channel $\sigma^R = 3 \sqrt{B+(0.05 B)^2}/ (\epsilon_s \mathcal{L})$, the Eq.~(\ref{eq:comb1}) can be simplified to
\begin{align}
\sigma^R(NN) = 1/ \sqrt{\L\frac{\text{Br}(\text{di-}h)}{\sigma^R(\text{di-}h)}\R^2+\L\frac{\text{Br}(\text{di-}W)}{\sigma^R(\text{di-}W)}\R^2+\L\frac{\text{Br}(h W)}{\sigma^R(h W)}\R^2}. \label{eq:comb2}
\end{align} 
($\sigma^R(\text{di-}h)$, $\sigma^R(\text{di-}W)$, $\sigma^R(h W)$) are already shown in the right panels of Fig.(\ref{wl14TeV},~\ref{hh:result},~\ref{wh:result}), respectively, and moreover the branching ratios almost stay constant for a sufficiently heavy RHN. With these, we then obtain the combined signal reaches in Fig.~\ref{comb:result}.

First we consider the $Z'$-mediated RHN pair production. In the figures we demonstrate the search reaches for several different values of $g'$. For a heavy $Z'$, the LHC searches for high-mass di-lepton resonances at 8 TeV~\cite{Chatrchyan:2012oaa,Aad:2014cka} yield the strongest bound on $Z'$~\cite{Guo:2015lxa}, e.g., they give $M_{Z'} \gtrsim 3450$ GeV for $g' =0.5$; such a constraint is indicated by the vertical line at the lower $M_{Z'}$ end in each contour of the search reach. From the left panel of Fig.~\ref{comb:result} we find that there is still a considerably wide region (with heavy $Z'$ of multi-TeV) which can be explored at the 14 TeV HL-LHC; RHN can be probed up to 2 TeV. At 100 TeV (the right panel), the search region can be substantially widened: fixing $g'=0.5$, RHN with mass 6 TeV is reachable for $M_{Z'}$ within 10-20 TeV.

Even though our simulations and analyses are based on the process of $Z'$ mediated RHN pair production, the results should be also applicable to other RHN pair production processes with similar kinematic properties, e.g., $g g \to H \to NN$. The loop-induced mode of MG5\_aMC@NLO~\cite{Hirschi:2015iia} is used to calculate the production cross section of this process, which is then mapped onto the $M_{X}-M_N$ plane in Fig.~\ref{comb:result}. When the mixing between the SM-like Higgs and the heavy Higgs is sizeable ($\sin \theta \gtrsim 0.2$), the HL-LHC is capable of reaching the parameter space of $M_\phi \lesssim 1.5$ TeV and $M_N \lesssim 500$ GeV. Obviously, the mass reaches of this case are significantly lower than those of the $Z'$ mediated case, because for the gluon initiated RHN production, its cross section dramatically drops in the high energy region owing to the PDF effect. However, the search of $H$-mediation is (partially) complementary to that of the $Z'$-mediation in the sense that the latter is not sensitive to the relatively light RHN, e.g, close to 200 GeV; as a reminder, searching for very heavy $M_{Z'}$ but light $M_N$ encounters problems like lepton overlapping and so on.

 \section{Conclusion}

In many UV completions of the type-I seesaw mechanism, RHNs participate in additional gauge or Yukawa interactions which contribute to RHN pair production 
via a $s$-channel resonance such as a vector boson $Z'$ or scalar boson $\phi$ which mixes with the SM Higgs boson $h$. Such a scenario provides a promising chance to probe heavy RHN without needing large mixing between RHN and 
light active neutrinos.

In this paper, we performed a model independent study of three signatures for the 
heavy RHN pair: trilepton from the di-$W$ channel, boosted di-Higgs plus \ET from the di-$h$ channel and the hybrid from the $hW$ channel. 
Our studies are specific to the 14 TeV LHC as well as the future 100 TeV $pp$-collider. 
For each signature, the search strategy is optimized on 4(5) benchmark points 
at the 14(100) TeV collider by using the BDT method.
Accordingly, 4(5) signal regions are defined. Those benchmark points with dramatically different kinematic properties are supposed to represent a large 
portion of the parameter space in their vicinities, and therefore those signal regions 
can be applied to other grids on the $M_{Z'}$-$M_{N}$ plane with $M_{Z'}>2M_N$ 
to draw a tentative global picture of RHN pair searches. Our studying shows that, 
for most grids trilepton always give the most sensitive signature, except in some 
corner of the plane which gives rise to the highly boosted RHN pair thus rendering 
the primary lepton too close to the hadronically decaying $W$. Then, the other signatures from the di-$h$ and $hW$ channels are complementary to the the trilepton signature.

We apply our searches to the benchmark model BLSM and find that the multi-TeV RHN can be probed at 14 TeV LHC. At the 100 TeV $pp$ collider, even the 
remarkable 10 TeV mass scale RHN could be probed, which enables us to cover most of the parameter space of low scale seesaw mechanism.

\section*{Acknowledgement}

We thank for the initial collaboration with Biswas Sanjoy.  This work was supported
in part by National Research Foundation of Korea (NRF) Research 
Grant NRF-2015R1A2A1A05001869 (PK), and by the NRF grant funded by the 
Korea government (MSIP) (No. 2009-0083526) through Korea Neutrino Research 
Center at Seoul National University (PK).

\appendix

\section{Benchmark model: local $B-L$ extended standard model (BLSM)}\label{spectra}


The BLSM is well motivated to understand neutrino physics since it naturally requires three RHNs to fulfill anomaly cancellation. It offers both resonances given in the simplified models Eq.~(\ref{EFT1},\ref{EFT2}). The particle content of the minimal BLSM is listed in Table.~\ref{particels}.
\begin{table}[htb]
\begin{center}
  \begin{tabular}{|c|c|c|c|c|c|c|c|c|}\hline
   & $q_L$ & $u_R$ & $d_R$ & $l_L$ & $e_R$ & ${N}_R$ & $H$ & $\Phi$ \\ \hline
  $SU(3)_c$ & 3 & 3 & 3 & 1 & 1 & 1 & 1 &1\\ \hline
  $SU(2)_L$ & 2 & 1 & 1 & 2 &1 & 1 &2 &1 \\ \hline
  $U(1)_Y$ & 1/6 & 2/3 & -1/3 & -1/2 & -1 & 0 & 1/2 & 0 \\ \hline
  $U(1)_{B-L}$  & 1/3 & 1/3 & 1/3 & -1 & -1 & -1 & 0 & 2 \\\hline
  \end{tabular}
  \caption{Field content and quantum numbers in the minimal BLSM.}
  \label{particels}.
\end{center}
 \end{table}

We briefly introduce the structure of this model. For a more detailed description, we refer to Ref.~\cite{Basso:2008iv}. In the first, the Higgs potential of the model is 
\begin{align}\label{potential}
V(\Phi,H)={m_1^2}|H|^2+m_2^2|\Phi|^2+\f{\ld_{2}}{4}|\Phi|^4+\f{\ld_{1}}{4}|H|^4+\f{\ld_{12}}{4}|\Phi|^2|H|^2,
\end{align} 
which is true for any local $U(1)$ extension, not only for $U(1)_{B-L}$. The Higgs field $\Phi$ develops a large VEV $v_X$ to break $B-L$, generating masses to $Z_{B-L}$ and RHNs:
\begin{align}
M_{Z_{B-L}}=2g_{B-L}\f{v_X}{\sqrt{2}},\quad M_N=\ld_N \f{v_X}{\sqrt{2}}. 
\end{align}
It is ready to solve the spectra of Higgs bosons, which obtain the mass squared  
\begin{align}\label{eq:mh1}
m_{H_{1,2}}^2=v^2\left[\f{\ld_1}{4}+\f{\ld_2 }{4}R^2 \mp\sqrt{\L\f{\ld_1}{4}-\f{\ld_2 }{4}R^2 \R^2+\f{\ld_{12}^2}{4}R^2}\right],
\end{align} 
and as well the mixing angle $\theta$ defined by $\tan2\theta=2\ld_{12} R/(\ld_1-\ld_2 R^2)$ with $R=v_X/v\gg1$ to guarantee the presence of a heavy Higgs boson $\phi\equiv H_2$ (with $h\equiv H_1$). In the small mixing limit and utilizing $\sin\theta\approx -\ld_{12}/(\ld_2 R)$ one can approximate\begin{align}\label{eq:mh2}
m_h^2\approx \f{\ld_1}{2} v^2-{\sin^2\theta}m_\phi^2,\quad m_{\phi}^2\approx \f{\ld_2}{2}R^2v^2.
\end{align}
Next we move to the gauge and Yukawa sectors. We do not want to list the complete terms in the Lagrangian but merely the relevant terms: 
\begin{align}
{\cal L}\supset   - {\cal Q}_f g_{B-L} \bar{f} \gamma_\mu f Z_{B-L}^\mu -\L y_N \bar{\ell}_L {N_R} \tilde{H} - \f{1}{2}\ld_N \overline{ (N_R)^c} {N_R} \Phi + h.c.\R,
\end{align}
where $f$ runs over all $B-L$ charged fermions, carrying charge ${\cal Q}_f$. With it one can calculate the decay widths of $ Z_{B-L}$ into fermions,
\begin{align}
\Gamma( Z_{B-L}\ra \bar ff)=\f{M_{ Z_{B-L}}}{12\pi}C_f({\cal Q}_{f}g_{B-L})^2\left(1+2\f{m_f^2}{M_{ Z_{B-L}}^2}\right)\sqrt{1-\f{4m_f^2}{M_{Z_{B-L}}^2}},
\end{align}
with $C_f$ the color factor. In the massless limit of fermions, the bracing ratio to three generations of RHN is $\lesssim 18\%$. While the branching ratio into a specific flavor is smaller than $6\%$. We restrict the discussions into the scenario where narrow width approximation (NWA) is hold, as yields the upper bound on the gauge coupling:
\begin{align}
\sum_f\Gamma( Z_{B-L}\ra \bar ff)\lesssim C_{\rm NWA} M_{ Z_{B-L}}\Rightarrow g_{B-L}\lesssim 0.49\L\f{ C_{\rm NWA} }{0.1}\R^{1/2}.
\end{align}
$C_{\rm NWA}$ is an artificial parameter, supposed to be sufficiently small to maintain NWA.



\section{Parameterization of the type-I seesaw mechanism}\label{mixing:U}

In this section we introduce the widely used parameterization of type-I seesaw mechanism Eq.~(\ref{seesaw}). Including three generations of RHNs, the masses and mixings can be described by the $6 \times 6$ mass matrix ${\cal M}$, which is diagonalized through the following unitary matrix $U$:
\begin{align}\label{}
U^T{\cal M} U=   \left(\begin{array}{cc}  m_{\nu}^{dia}   &   \\  &  M_{N}^{dia}
\end{array}\right)
~~ {\rm with} ~~ U=\left(\begin{array}{cc} U_{PMNS} & U_{\nu N} \\ U_{N\nu} & 1
\end{array}\right),
\end{align} 
where $m_{\nu}^{dia} ={\rm diag}(m_{\nu_1},m_{\nu_2},m_{\nu_3})$ and  $M_{N}^{dia} ={\rm diag}(M_{N_1},M_{N_1},M_{N_1})$. The charged leptons are assumed to be already in the mass basis. At leading order in $m_D/M_N$ (with $m_D=y_N v/\sqrt{2}$ the Dirac mass matrix), the mixing matrices are given by
\begin{align}\label{}
U_{\nu N}=m_{D}^\dagger (M_{N}^{dia})^{-1},\quad U_{ N\nu}=-(M_{N}^{dia})^{-1} m_D U_{PMNS},
\end{align} 
It is convenient to introduce the induced light neutrino mass matrix $m_\nu\equiv -U_{\nu N} m_D^*$, which is diagnoalized by the PMNS matrix, leading to $m_{\nu}^{dia}=U_{PMNS}^\dagger m_\nu U_{PMNS}^*$. Using these relations, one can derive 
the following equation for $U_{\nu N}$, 
\begin{align}\label{}
U_{PMNS} m_{\nu}^{dia} U_{PMNS}^T=-U_{\nu N}  M_{N}^{dia}U_{\nu N}^T.
\end{align} 
Considering the $ii-$element of both sides of above equation, one gets the relation::
\begin{align}\label{}
-\sum_\alpha (U_{\nu N})_{i\alpha}^2 M_{N_\alpha}=\sum_j(U_{PMNS})_{ij}^2m_{\nu_j}.
\end{align} 
For the degenerate RHNs, we have 
\[
U_i\equiv \sum_\alpha (U_{\nu N})_{i\alpha}^2=\sum_j(U_{PMNS})_{ij}^2m_{\nu_j}/M_N.
\]


\end{document}